\documentclass[fleqn,usenatbib]{mnras}
\usepackage[T1]{fontenc}
\usepackage{ae,aecompl}
\usepackage{amsmath}
\usepackage{graphicx}
\usepackage{epsfig}
\usepackage{amssymb,txfonts}
\usepackage{xspace,hyperref}
\usepackage[T1]{fontenc}
\usepackage{ae,aecompl}

\newcommand{\msun}{{\rm M}_{\sun}}
\newcommand{\thcomp}{\texttt{thComp}\xspace}
\newcommand{\compton}{\texttt{compton}\xspace}
\newcommand{\compps}{\texttt{compps}\xspace}
\newcommand{\comptt}{\texttt{comptt}\xspace}
\newcommand{\xspec}{\textsc{xspec}\xspace}

\newbox\grsign \setbox\grsign=\hbox{$>$} \newdimen\grdimen \grdimen=\ht\grsign
\newbox\simpropbox
\setbox\simpropbox=\hbox{\raise.5ex\hbox{$\propto$}\llap
 {\lower.5ex\hbox{$\sim$}}}\ht2=\grdimen\dp2=0pt

\title[Compton scattering]{Spectral and temporal properties of Compton scattering by mildly relativistic thermal electrons}

\author[A. A. Zdziarski et al.]
{Andrzej A.~Zdziarski,$^{1}$\thanks{E-mail: aaz@camk.edu.pl (AAZ); mitsza@camk.edu.pl (MS); juri.poutanen@utu.fi (JP)} Micha{\l} Szanecki,$^1$\footnotemark[1] Juri Poutanen,$^{2,3,4}$\footnotemark[1]
Marek Gierli\'nski$^{5}$\newauthor and Pawe{\l} Biernacki$^1$\\
$^{1}$Nicolaus Copernicus Astronomical Center, Polish Academy of Sciences, Bartycka 18, 00-716 Warszawa, Poland\\
$^{2}$Tuorla Observatory, Department of Physics and Astronomy, FI-20014 University of Turku, Finland\\
$^{3}$Space Research Institute, Russian Academy of Sciences, Profsoyuznaya 84/32, 117997 Moscow, Russia\\
$^{4}$Nordita, KTH Royal Institute of Technology and Stockholm University, Roslagstullsbacken 23, SE-10691 Stockholm, Sweden\\
$^{5}$School of Life Sciences, University of Dundee, Dundee, UK}

 \date{Accepted 2020 January 14. Received 2019 December 22; in original form 2019 October 10}
 
\pubyear{2019}

\begin{document}
\defcitealias{st80}{ST80}
\label{firstpage}
\pagerange{\pageref{firstpage}--\pageref{lastpage}}
\maketitle

\begin{abstract}
We have obtained new solutions and methods for the process of thermal Comptonization. We modify the solution to the kinetic equation of Sunyaev \& Titarchuk to allow its application up to mildly relativistic electron temperatures and optical depths $\ga{1}$. The solution can be used for spectral fitting of X-ray spectra from astrophysical sources. We also have developed an accurate Monte Carlo method for calculating spectra and timing properties of thermal-Comptonization sources. The accuracy of our kinetic-equation solution is verified by comparison with the Monte Carlo results. We also compare our results with those of other publicly available methods. Furthermore, based on our Monte Carlo code, we present distributions of the photon emission times and the evolution of the average photon energy for both up and down-scattering.
\end{abstract}

\begin{keywords}
accretion, accretion discs -- radiation mechanisms: thermal -- radiative transfer -- scattering -- galaxies: active -- X-rays: binaries
\end{keywords}

\section{Introduction}
\label{intro}

Compton scattering is a major physical process in many different types of astrophysical sources. In accreting systems, the electrons often appear thermal at mildly relativistic temperatures, $kT_{\rm e}$. Then, they can Compton upscatter some soft seed photons, e.g., blackbody-like emission of optically-thick accretion discs \citep{ss73} or thermal synchrotron emission (e.g., \citealt{wardzinski00}) up to the hard X-ray range. The latter becomes inefficient at $kT_{\rm e}\la 100$\,keV. However, even a tiny non-thermal tail beyond the Maxwellian electron distribution increases the synchrotron emission significantly, which can be important in many astrophysical sources, see, e.g., \citet{wardzinski01}, \citet{poutanen09}, \citet{malzac09}, \citet*{veledina11}, \citet*{veledina13}.

In particular, thermal Comptonization appears to be responsible for most of the X-ray emission of hot coronae in accreting X-ray binaries in the hard spectral state, see, e.g., \citet{g97}, \citet{zdziarski98}, \citet{wardzinski02}, \citet{zg04}, \citet*{dgk07}, \citet*{burke17} and references therein. Similarly, high-energy cutoffs indicating the presence of thermal Comptonization are commonly observed in unbeamed AGNs, e.g., \citet{madejski95}, \citet{gondek96}, \citet*{zpj00}, \citet{zg01}, \citet{malizia08}, \citet{lubinski10,lubinski16}, \citet{brenneman14}, \citet{ballantyne14}, \citet{marinucci14}, \citet{balokovic15}, \citet{fabian15}, \citet{tortosa18}.

In the limit of large Thomson optical depth, $\tau \gg 1$, and non-relativistic electron temperatures, $kT_{\rm e}\ll m_{\rm e} c^2$ (where $m_{\rm e}c^2\approx 511$\, keV is the electron rest energy), the classical paper of \citet{st80} (hereafter \citetalias{st80}) provides analytical solutions for the spectra from thermal Comptonization, based on applying the kinetic (Fokker-Planck) equation of \citet{kompaneets56, kompaneets57} and treating photon escape as diffusive. Given that their solution is non-relativistic and based on the diffusion approximation, those spectra become inaccurate for either low optical depths, $\tau\la 3$ (as noted in \citetalias{st80}) or at high temperatures, $kT_{\rm e}\ga 100$ keV (or both). Those parameter ranges overlap with those found in accreting sources, see the references above.

\citet{cooper71} proposed to introduce a kinetic equation with coefficients giving exact relativistic energy exchange and dispersion coefficients and guaranteed photon conservation and Wien equilibrium. However, his solution suffered from some errors. Then, the Kompaneets equation with some of the coefficients of \citet{cooper71} was applied to Comptonization in the mildly relativistic and moderately optically thick regimes by \citet*{sle76}, \citet{lz87} and \citet*{zjm96}. The results of \citet{zjm96} were implemented into the \xspec \citep{arnaud96} suite for spectral fitting by \citet*{zycki99} as the fitting routine \texttt{Nthcomp}. The formulation of \citet{zjm96} also allowed to parametrize the spectrum in terms of the low-energy, $E\ll kT_{\rm e}$, power-law index of the spectrum, $\Gamma$ (where the energy flux $F(E)\propto E^{1-\Gamma}$). However, as shown in detail by \citet*{nsz19}, \texttt{Nthcomp} provides quite inaccurate results already at midly relativistic temperatures and $\tau$ of a few. In particular, the high-energy cutoffs of \texttt{Nthcomp} are too low. 

Different sets of corrections to the solution of \citetalias{st80} were used by \citet{titarchuk94}, \citet{titarchuk95} and \citet{hua95}. In particular, \citet{hua95} derived a formula for the spectrum covering both the non-relativistic and sub-relativistic temperature regimes and both low and high optical depths. Their results were based on comparison with Monte Carlo results and were implemented by them as the fitting routine \texttt{comptt} within \xspec. However, their solution still shows relatively large deviations from the exact Monte Carlo results, especially for $\tau\la 2$ and $kT_{\rm e}\ga 100$ keV, as shown in their figs.\ 3--6.

Another approach to the problem was followed by \citet{ps96}. They used an iterative scattering method (ISM) which solves the integral kinetic equation exactly. This can, in principle, provide accurate solutions for any temperature and optical depth. Their method was implemented as the fitting routine \texttt{compps} (in \xspec). However, that implementation requires a very large number of iterations for large optical depths, in particular at low temperatures. Also, the spectra are not parameterized by $\Gamma$.

Yet another approach is to use the Monte Carlo method. This was first done by \citet*{pozdnyakov76, pozdnyakov77, pozdnyakov83}. This method was also developed by \citet{gw84} and \citet{g00}. However, the Monte Carlo method is not suitable for directly spectral fitting of X-ray spectra due to the random scatter of the model spectra intrinsic to the method. A way to deal with this problem is to fit a table model generated from a large set of Monte Carlo spectra, but unless a very large set is used, such a model covers a limited range of the parameters.

Accurate determination of physical parameters of the X-ray sources in accreting X-ray binaries and AGNs is of major importance. In particular, correct measurements of the electron temperature are crucial for determination of the role of e$^\pm$ pair production in those sources (e.g., \citealt{z85,stern95,fabian15}). Also, correct measurements of the optical depths of those sources are important for determination of the location and physics of those sources, see, e.g., \citet{middei19}. Given the above issues and the problems with the existing methods discussed above, we have embarked on finding new accurate solutions to the Comptonization problems.

Specifically, we have developed the code of \citet{gw84} and \citet{g00}, and made its resulting version, \compton\footnote{\url{http://users.camk.edu.pl/mitsza/compton/}}, publicly available. The current code gives both the spectra for a variety of geometries and spatial distributions of seed photons and the distributions of the time spent by a photon in the source, which represent Green's function, $G$, for the problem. We show the form of $G$ for a number of different source parameters. The code also calculates the time lags due to Compton scattering, which we show here. Furthermore, it can take into account absorption of photons, which, however, is not treated in the present paper. For the sake of simplicity, we present here results of \compton only for the spherical geometry.

We have then used our Monte Carlo results to find a set of corrections to the kinetic equation of \citetalias{st80}, allowing it to be applied to spherical geometry with $\tau \ga 1.6$ and $kT_{\rm e}\la 300$\,keV. Our correction factors are based mostly on the comparison with the Monte Carlo results, and are phenomenological. Also, we include the unscattered part of the seed photon spectrum in the observed spectrum. Furthermore, our new public fitting routine, \thcomp\footnote{\url{http://users.camk.edu.pl/mitsza/thcomp/}, \url{http://github.com/HEASARC/xspec_localmodels/}} in \xspec, is provided as a convolution function, and thus it can be used with any form of seed photons. We test its accuracy for both the problem of Compton upscattering of soft photons by mildly relativistic electrons and Compton down-scattering of photons at $E\la m_{\rm e}c^2$ by cold electrons. We also use our Monte Carlo results of \compton to determine the detailed limits of the applicability of \thcomp, and to compare with the results of \compps and \comptt. 

\section{Kinetic equations for Compton scattering}
\label{kinetic}

To place our results in a context, we review first some of the previous results on the kinetic equation for Compton scattering. This integral kinetic equation gives a complete description of interactions between photons and electrons via Compton scattering. It can be written in some specific form, for example, using the Fokker-Plank approximation, and it then becomes a differential equation -- a diffusion equation in frequency. If the equation is applied to spatial transport, it can also be simplified and written as a diffusion equation in spatial coordinates (as done by \citetalias{st80}). 

Let us first consider only evolution of the radiation spectrum with frequency in homogeneous medium, i.e.\ we omit the terms containing spatial derivatives. Neglecting, for the sake of simplicity, the stimulated scattering term, the Fokker-Planck equation for this problem is \citep[e.g.][]{vurm09}
\begin{equation}
\frac{\partial n}{\partial t}= n_{\rm e}\sigma_{\rm T} c \frac{\partial}{\partial\epsilon} \left\{\frac{\partial}{\partial\epsilon} \left[\frac{\langle(\epsilon_1- \epsilon)^2\rangle}{2} n\right] -\langle\epsilon_1- \epsilon\rangle n \right\},\label{fk}
\end{equation}
where $n$ is photon density per unit volume and per dimensionless photon energy is $\epsilon\equiv E/m_{\rm e}c^2$, $\sigma_{\rm T}$ is the Thomson cross section, $n_{\rm e}$ is the electron density, $\epsilon_1$ is the photon energy after a scattering, $\langle\epsilon_1- \epsilon\rangle$ and $\langle(\epsilon_1- \epsilon)^2\rangle$ are the average energy shift and the average square of the energy shift, respectively, and $\epsilon_1- \epsilon\ll \epsilon$ is assumed. For Compton scattering by thermal electrons in the Thomson limit \citep[e.g][]{nagirner94},
\begin{equation}
\langle\epsilon_1- \epsilon\rangle=(4\theta-\epsilon)\epsilon,\quad \langle(\epsilon_1- \epsilon)^2\rangle =2\theta\epsilon^2,
\label{coeffNR}
\end{equation}
where $\theta\equiv k T_{\rm e}/m_{\rm e}c^2$ is dimensionless electron temperature. Substituting these coefficients in equation (\ref{fk}) gives us the equation of \citet{kompaneets56, kompaneets57}
\begin{equation}
\frac{\partial n}{\partial t}= n_{\rm e}\sigma_{\rm T} c \frac{\partial}{\partial\epsilon} \left[\epsilon^4 \left(\theta \frac{\partial}{\partial\epsilon} \frac{n}{\epsilon^2} +\frac{n}{\epsilon^2} \right)\right].\label{k56}
\end{equation}
The equation is valid for $\theta\ll 1$, $\epsilon \ll 1$. However, it was noted by \citet*{ross78} that this equation fails for $\theta\ll \epsilon$. Namely, at $\theta=0$ equation (\ref{k56}) becomes a first-order differential equation, which only shifts the photon energy down neglecting the associated dispersion. If the zero-temperature dispersion term is included, the equation becomes
\begin{equation}
\frac{\partial n}{\partial t}= n_{\rm e}\sigma_{\rm T} c \frac{\partial}{\partial\epsilon} \left\{ \epsilon^4 \left[ \left( \theta+\frac{7}{10}\epsilon^2\right) \frac{\partial}{\partial\epsilon} \frac{n}{\epsilon^2} +\frac{n}{\epsilon^2} \right] \right\} .\label{ross}
\end{equation}
This down-scattering problem was further studied by \citet{illarionov79}, who confirmed its correctness in application to Compton down-scattering of photons with $\epsilon\ll 1$ by cold electrons.

\citet{kershaw86}, \citet{prasad86, prasad88} and \citet{shestakov88} derived a fully relativistic kinetic equation, with the coefficients valid at any value of the electron temperature. This was further developed by \citet{nagirner94}, who derived polarization redistribution matrices. However, the form of the coefficients in those equations is relatively complex. 

Iterative relativistic corrections to the Kompaneets equation (including the induced scattering) were derived by \citet{challinor98}. They presented a formalism allowing the corrections to be included up to any desired order, and gave the explicit expressions up to the second order in $\theta$ and $\epsilon$. The expansion series given by their equation (16) contains, in fact, the $(7/10)\epsilon^2$ term included by \citet{ross78}, and their results are valid for both up and down-scattering. Their results were extended in further studies by \citet*{itoh98}, \citet{brown12} and \citet{nozawa15}, who confirmed the correctness of the results of \citet{challinor98}. However, their equation is a fourth-order partial differential equation and the obtained series is only asymptotic. Thus, its application to the problems of Comptonization by mildly relativistic thermal plasma and of Compton down-scattering of radiation with $\epsilon\sim 1$ by cold plasma is not practical. On the other hand, the modification of the Kompaneets equation proposed by \citet{liu04} appears incorrect. Then, \citet{vurm09} treated Compton scattering using either integral or differential form depending on the value of the fractional photon energy change.

\citet{challinor98}, \citet{itoh98}, \citet{brown12} and \citet{nozawa15} also applied their results to the Sunyaev-Zeldovich effect. Without relativistic corrections, the Kompaneets equation (including the stimulated scattering term) only allows us to determine the Compton parameter, $y\equiv 4\theta\tau $, of the medium (hot gas with $kT_{\rm e}\sim 10$\,keV in a cluster of galaxies) upscattering the cosmic microwave background radiation \citep{zeldovich69}. However, with those corrections, both the temperature and the optical depth of the cluster gas can be determined. 

\section{The phenomenological kinetic equation}
\label{thcomp}

We would like now to consider not only evolution of the spectrum in frequency but also the effects related to the finite size of the sources. We simplify the treatment of radiative transfer by using escape probability approximation. In non-relativistic limit, the corresponding kinetic equation is given in \citetalias{st80}. We use the same form, but make some phenomenological modifications to include both relativistic corrections as well as corrections due to the optical depth of the system being relatively low. In steady-state, our equation is as follows,
\begin{equation}
\dot n_0(\epsilon) +\dot n_{\rm C}(\epsilon)+\dot n_{\rm esc}(\epsilon)=0. \label{eq}
\end{equation}
Here $n$ is now the photon spatial density averaged over the volume of the source, $\tau =n_{\rm e}\sigma_{\rm T} R$ is the Thomson optical depth, $R$ is a characteristic source size, and 
\begin{align} 
&\dot n_{\rm C}(\epsilon)=\frac{c}{R}\tau \frac{{\rm d}}{{{\rm d}\epsilon}}\left\{\omega
(\epsilon)\epsilon^4 \left[\theta\xi(\theta) \frac{\rm d}{{\rm d}\epsilon} \frac{n(\epsilon)}{\epsilon^2} +\frac{n(\epsilon)}{\epsilon^2} \right]\right\},\label{nc}\\
&\dot n_{\rm esc}(\epsilon)= -\frac{c}{R} \frac{n(\epsilon)\tau }{\bar u(\tau ,\theta,\epsilon)}, \label{nescape}
\end{align} 
are the rates of the Comptonization and photon escape, respectively. The first term, $\dot n_0$, is the rate of the seed photon production. Then, $\omega$ and $\xi$ are relativistic corrections to the photon energy shift, and $\bar u$ is the average number of scatterings, see below. We solve the above second-order ordinary differential equation numerically.

We note that the above equation neglects the stimulated scattering factor $(1+{\cal N})$, where ${\cal N}$ is the photon occupation number in the phase space. For a blackbody photon field,
\begin{equation}
{\cal N}=\frac{1}{\exp(\epsilon/\theta)-1}. 
\label{occupation}
\end{equation}
In that case, ${\cal N}> 1$ for $\epsilon< \theta\ln 2$. In accreting sources, the main blackbody-like photon source is an accretion disc, in which case the blackbody is usually significantly diluted, reducing ${\cal N}$. Furthermore, Compton upscattering by energetic electrons is strongly dominated by the part of the spectrum containing most of the energy, which is $\epsilon\ga \theta$. Therefore, neglecting ${\cal N}$ is a very good approximation in most cases of interest.

We then attempt to find the form of the correction factors, $\omega$ and $\xi_{\rm e}$, and of the average scattering number, $\bar u$, by comparing results of the solution of equations (\ref{eq}--\ref{nescape}) to those obtained using the Comptonization Monte Carlo method as implemented in the code \compton (described in Section \ref{intro}). We choose the spherical geometry, and assume a uniform electron density and the spatial density of sources of seed photons $\propto \tau'^{-1}\sin(\upi \tau'/\tau )$, where $\tau'$ is the optical depth measured from the centre. In this case, the distribution of the photon escape time at $\tau \gg 1$ is an eigenfunction of the diffusion equation, and it has a simple form, see equation (9) in \citetalias{st80}. This case is intermediate between the sources of photons uniformly distributed over the sphere and concentrated at its centre.

We find we achieve good agreement between the Monte Carlo results and the solutions of equations (\ref{eq}--\ref{nescape}) for
\begin{align}
&\xi(\theta)=1+\theta+3\theta^2,
\label{xi_theta}\\
&\omega(\epsilon)= (1+4.6\epsilon +1.1\epsilon^2)^{-1},
\label{cooper}
\end{align}
where former was fitted to the average fractional shift due to upscattering by thermal electrons increasing with temperature in the mildly relativistic regime and the latter is the fit to the mean fractional energy shift by cold electrons of \citet{cooper71}. The average number of scattering is modified by us with respect to the non-relativistic, optically-thick value \citepalias{st80} as follows,
\begin{align}
&\bar u(\tau ,\theta,\epsilon)=\tau \left[a_1(\theta)+a_2(\theta)\tau_{\rm C}(\epsilon) g(\epsilon)\right],\label{ubar}\\
&a_1(\theta)=\frac{1.2}{1+\theta+5\theta^2},\quad a_2(\theta)=\frac{0.25}{1+\theta+3\theta^2},\label{an}\\
&g(\epsilon)=\begin{cases}1,&\epsilon\leq 0.1;\cr 
(2-\epsilon)/1.9, &0.1<\epsilon<2;\cr 0\;,&\epsilon\geq 2.\cr \end{cases} \label{escape}.
\end{align}
Here, $g(\epsilon)$ accounts in an approximate way for the Klein-Nishina reduction of the scattering efficiency at mildly relativistic photon energies, and it is defined slightly different from \citet{lz87}, and $\tau_{\rm C}=n_{\rm e}\sigma_{\rm C} R$ and $\sigma_{\rm C}$ are the Compton optical depth and the total cross section (for cold electrons; e.g., \citealt{rl79}), respectively. In the limit of $\theta\ll 1$, $\tau \gg 1$, the above $\bar u$ is close to the corresponding solution of \citetalias{st80}, $\bar u=3(\tau ^3+2/3)^2/\upi^2$. Our phenomenological modifications (based on the Monte Carlo results) with respect to that result give a reduction of the average number of scatterings in the mildly relativistic regime.

We notice that our equations do not contain any zero-temperature dispersion term, which would be analogous to the $(7/10)\epsilon^2$ term in equation (\ref{ross}). This could limit their applicability to cases with low-energy electrons scattering photons with much higher energies. However, we have found that in spite of that down-scattering of broad continua gives correct results even for very low temperatures, see Section \ref{down}. Still, our equations would not give correct results for down-scattering by cold electrons of lines (cf.\ \citealt{illarionov79}). 

Also, in the limit of large $\tau$, the Wien equilibrium in our equations is established not at $\theta$ but at a higher temperature, $\xi(\theta)\theta$, which can be seen by setting $\dot n_{\rm C}=0$ in equation (\ref{nc}). While the correct equilibrium is still established at $\theta\ll 1$, our equations should, therefore, not be applied to cases with large $\tau$ and relativistic temperatures. However, such conditions do not seem to occur in accretion flows in X-ray binaries and AGNs. 

\begin{figure*}
\centerline{\includegraphics[width=7.7cm]{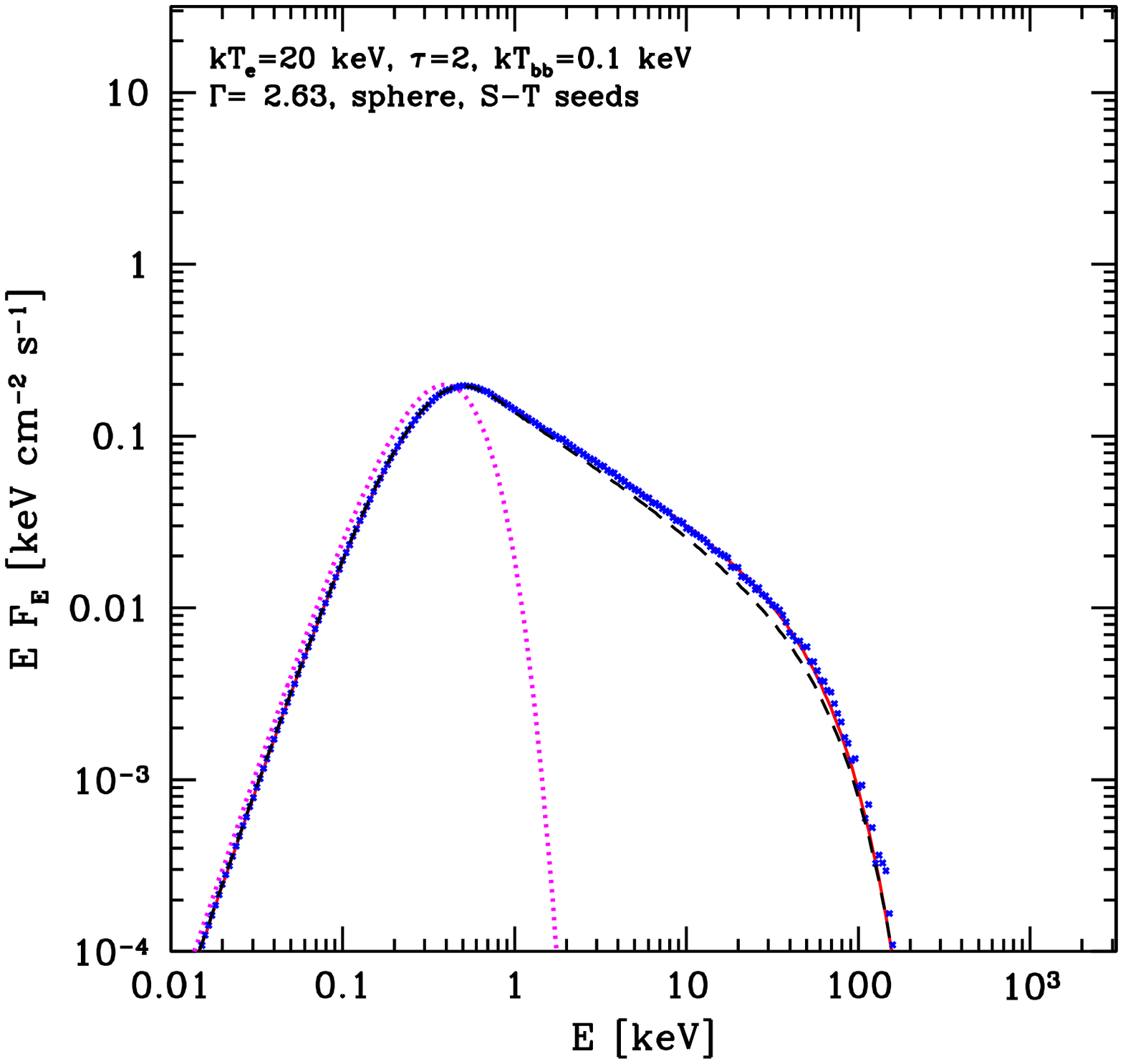}
\includegraphics[width=7.7cm]{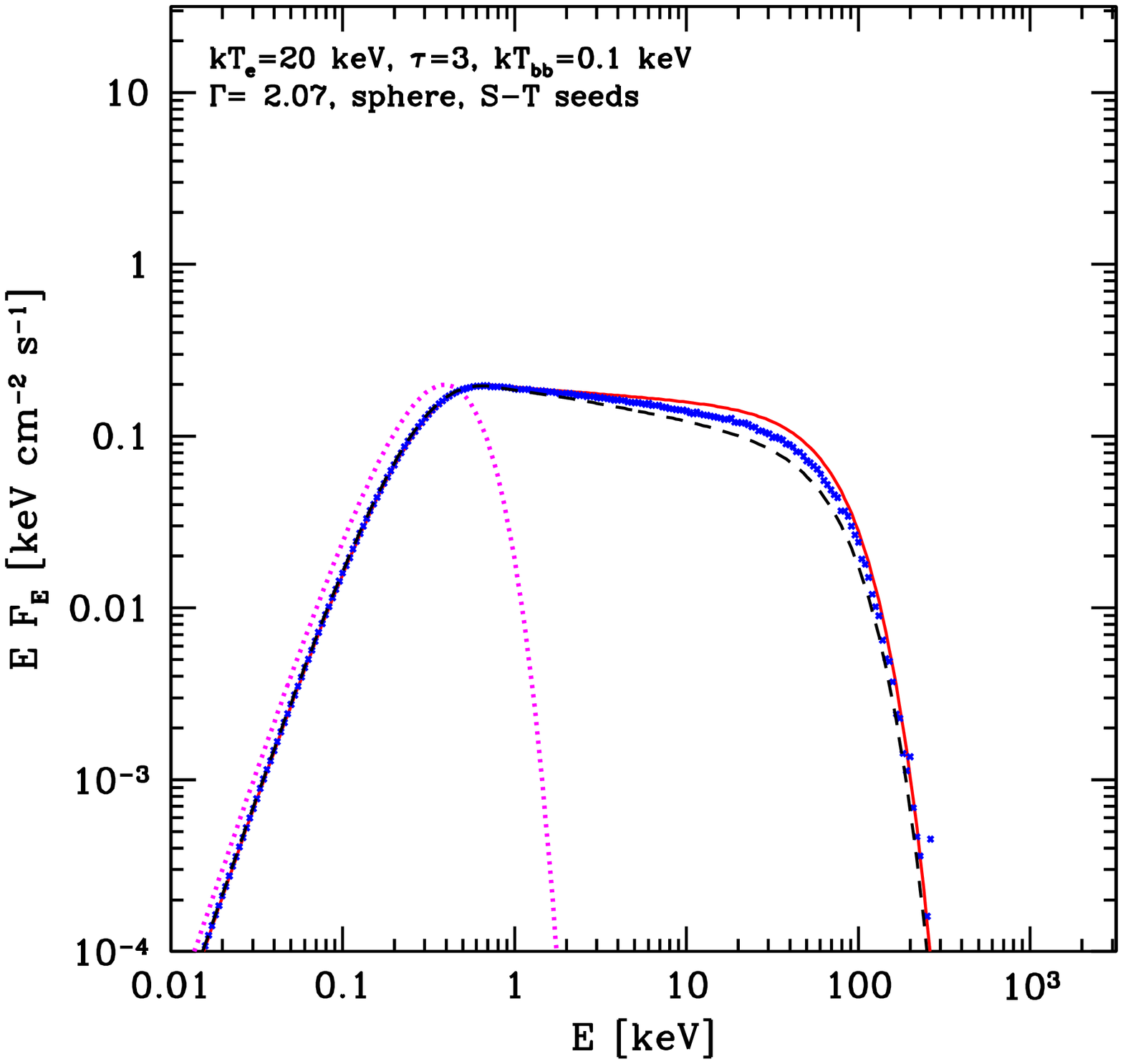}}
\centerline{\includegraphics[width=7.7cm]{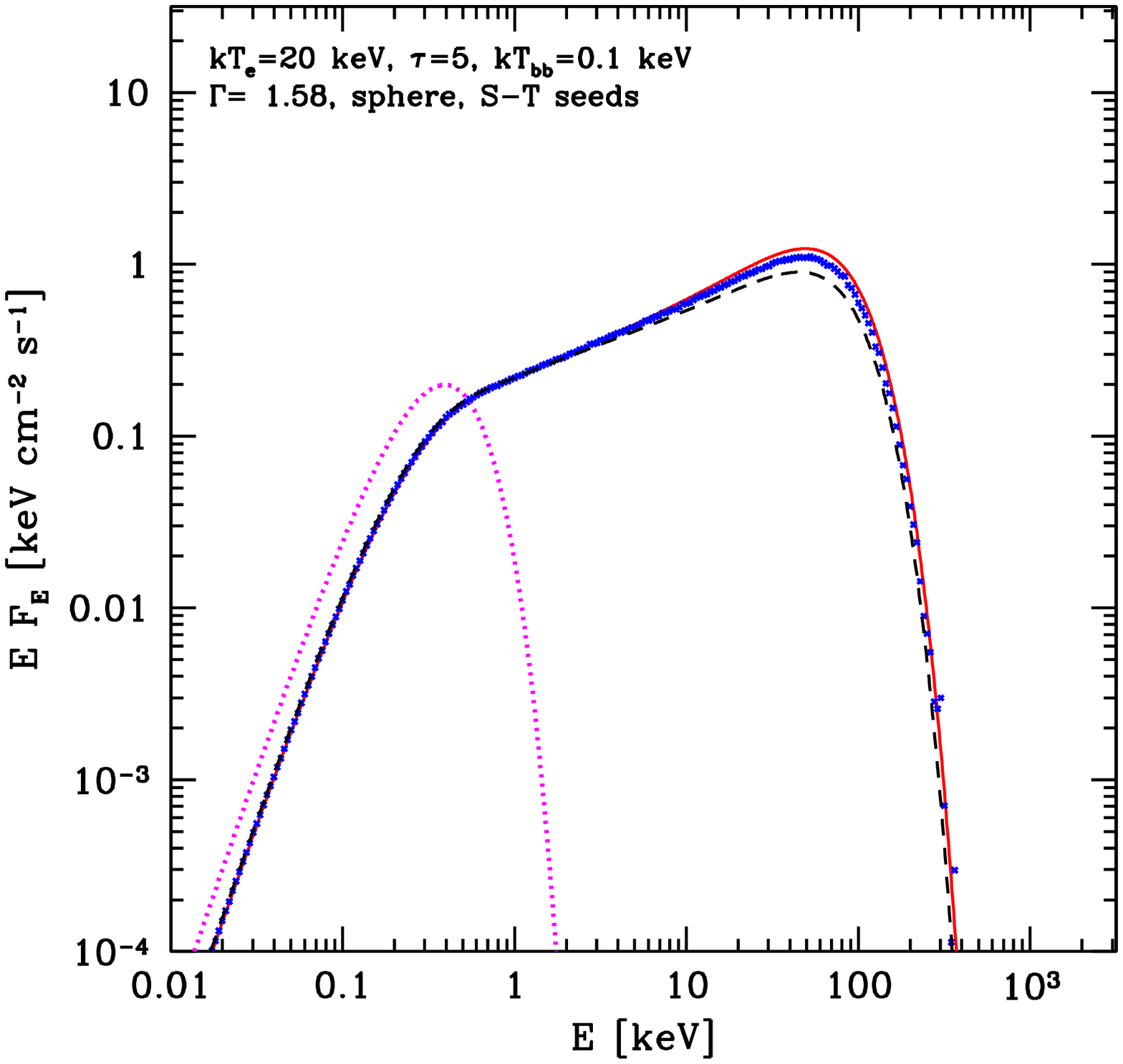} 
\includegraphics[width=7.7cm]{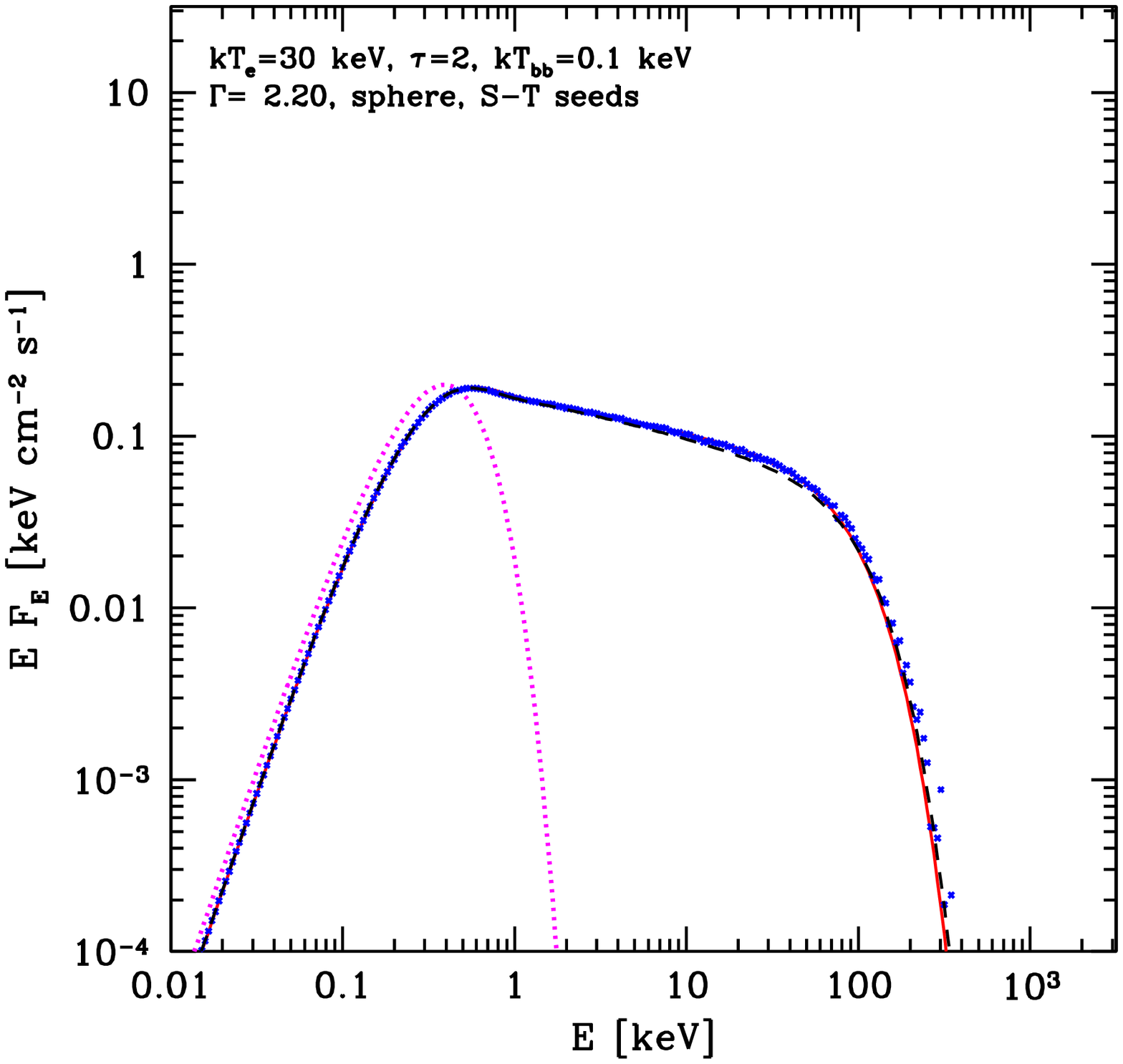}}
\centerline{\includegraphics[width=7.7cm]{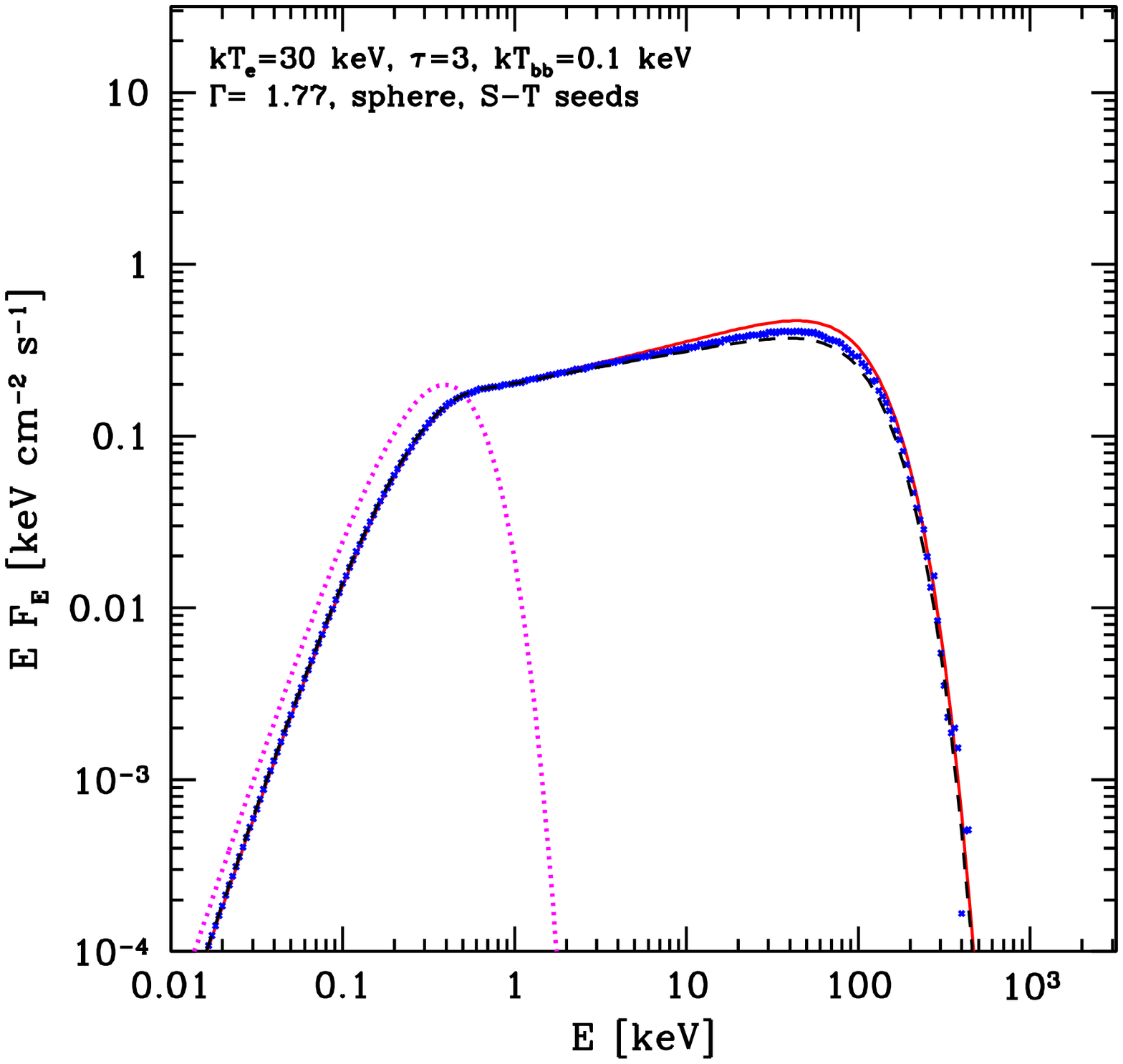}
\includegraphics[width=7.7cm]{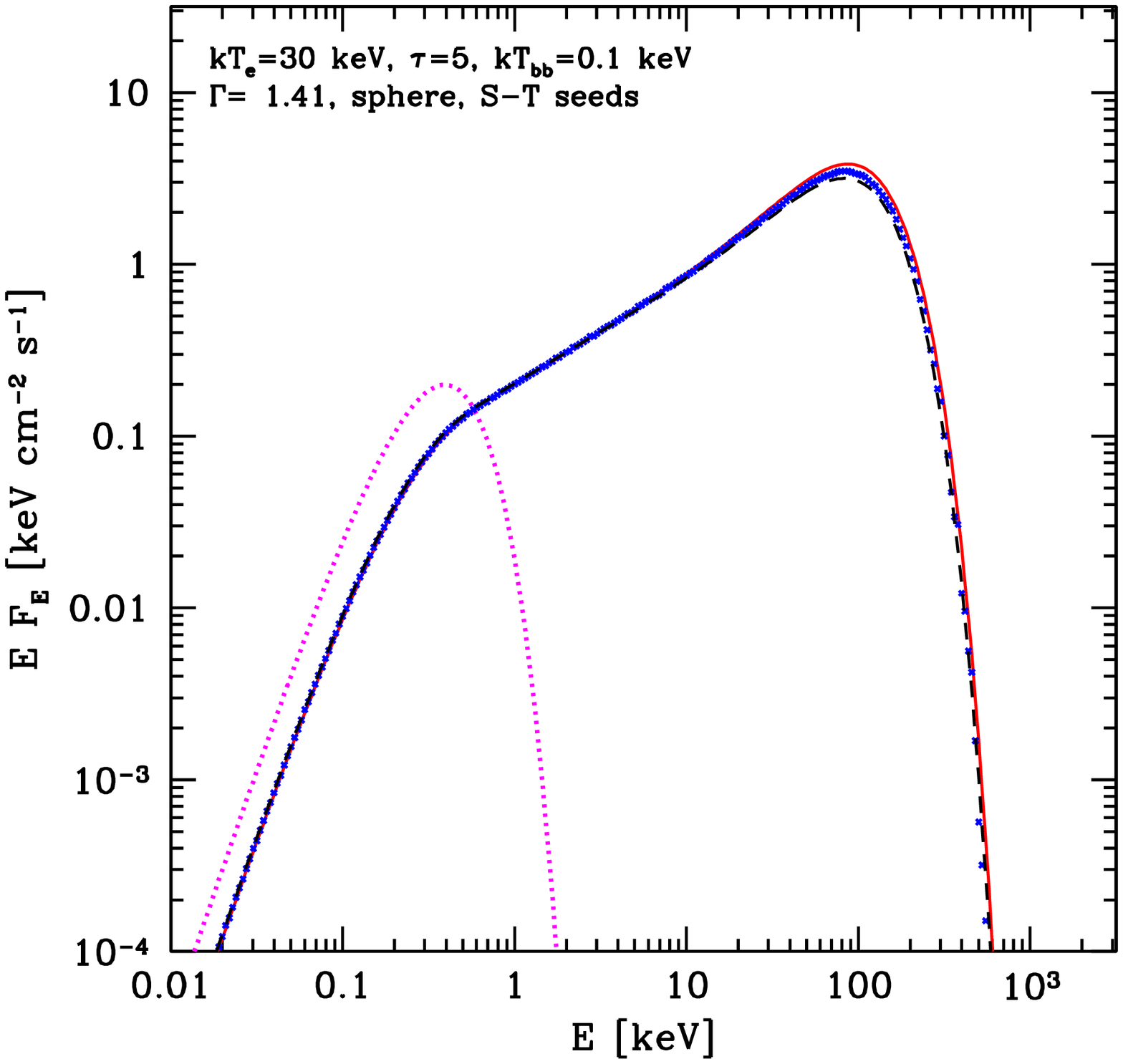}} 
\caption{Comparison of thermal Comptonization spectra calculated using the Monte Carlo method, \compton (blue points), the solution of our kinetic equation, \thcomp (red solid curves), and the ISM, \compps (black dashed curves), for $\tau \geq 2$ and $kT_{\rm e}\leq 150$ keV. The source is a homogeneous electron sphere with sources of blackbody seed photons (shown by magenta dotted curves) with a sinusoidal spatial distribution $\propto \tau'^{-1}\sin(\upi \tau'/\tau )$ (\citetalias{st80}). The values of $kT_{\rm e}$, $\tau$, $kT_{\rm bb}$, and the resulting $\Gamma$ are as given on the plots. In the shown range, there is a good agreement between \thcomp and the Monte Carlo results. The public version of the \compps code is also highly accurate, except for the cases with $kT_{\rm e}=20$ keV, where it somewhat underestimates the actual spectrum. The photon number is normalized to unity.
} \label{tau2_5}
\end{figure*}

\setcounter{figure}{0}
\begin{figure*}
\centerline{\includegraphics[width=7.7cm]{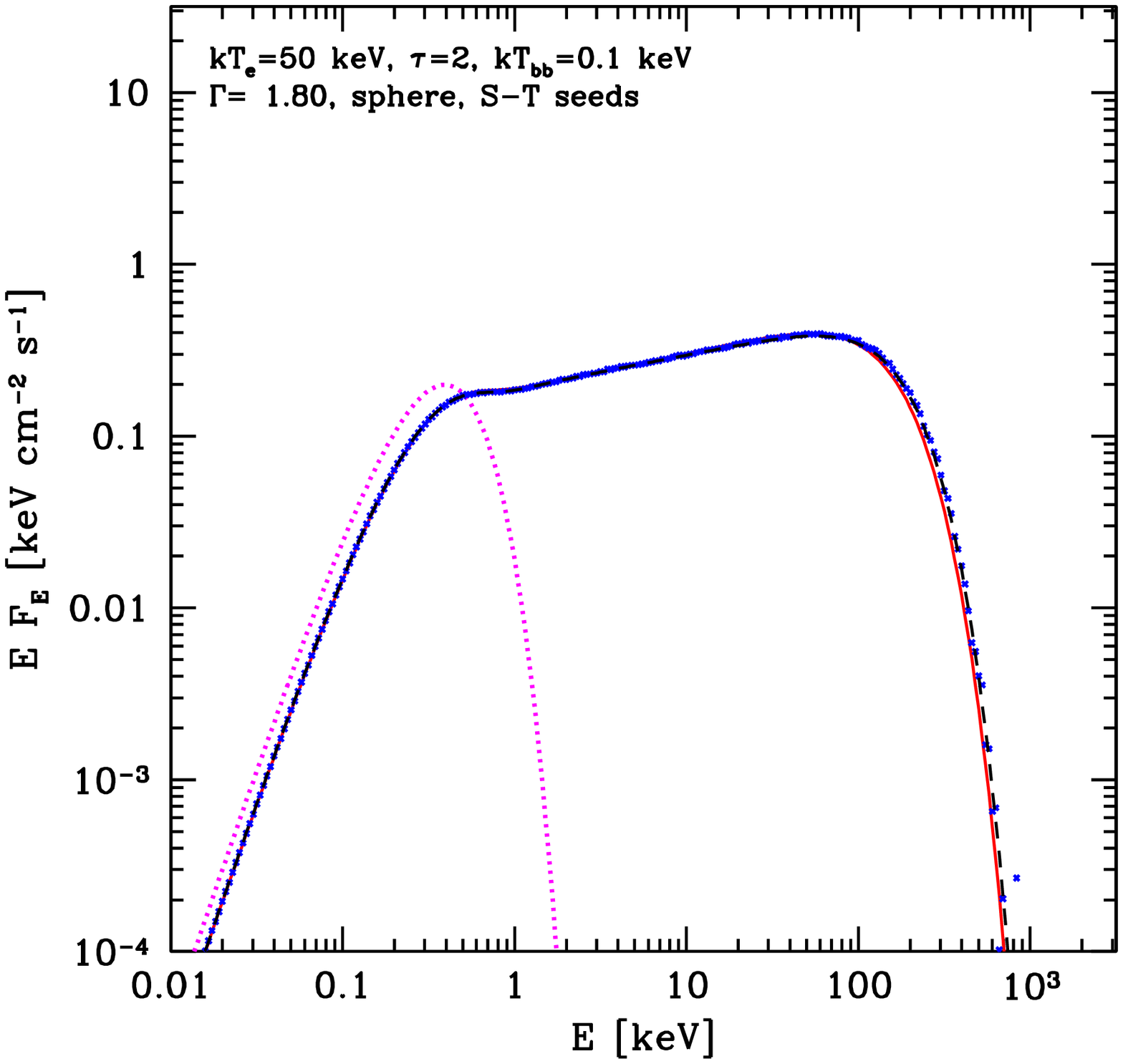}
\includegraphics[width=7.7cm]{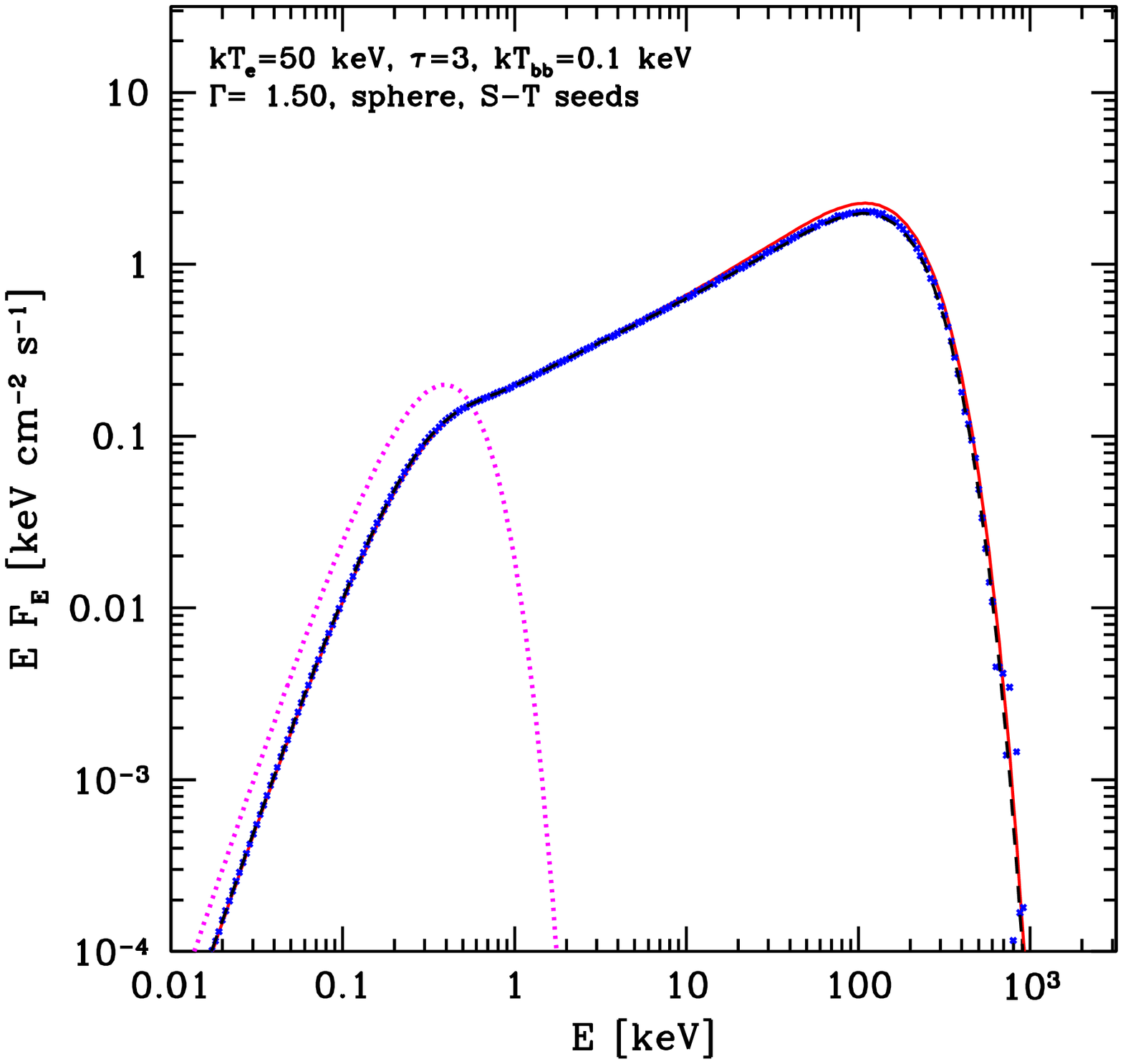}}
\centerline{\includegraphics[width=7.7cm]{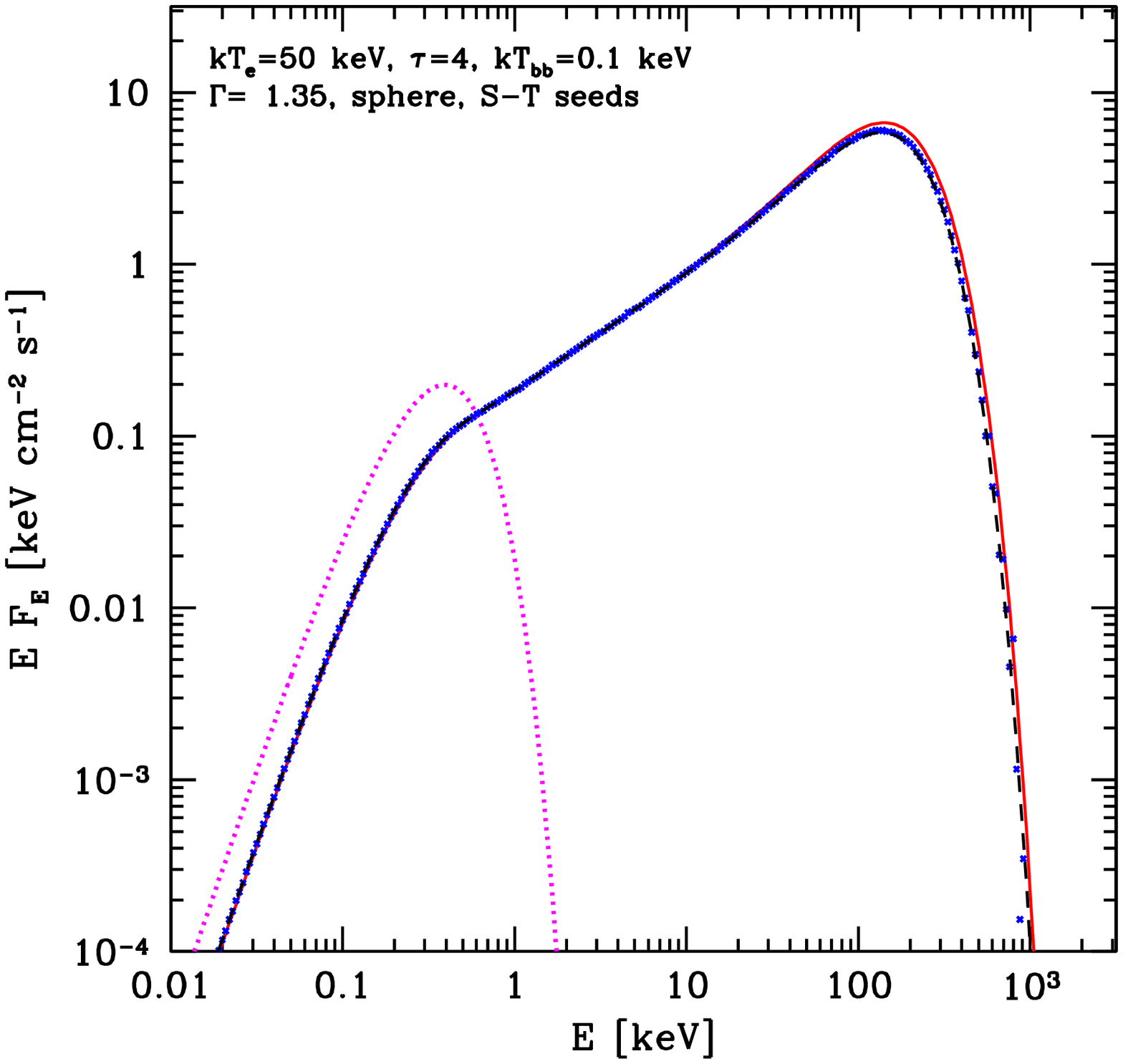} 
\includegraphics[width=7.7cm]{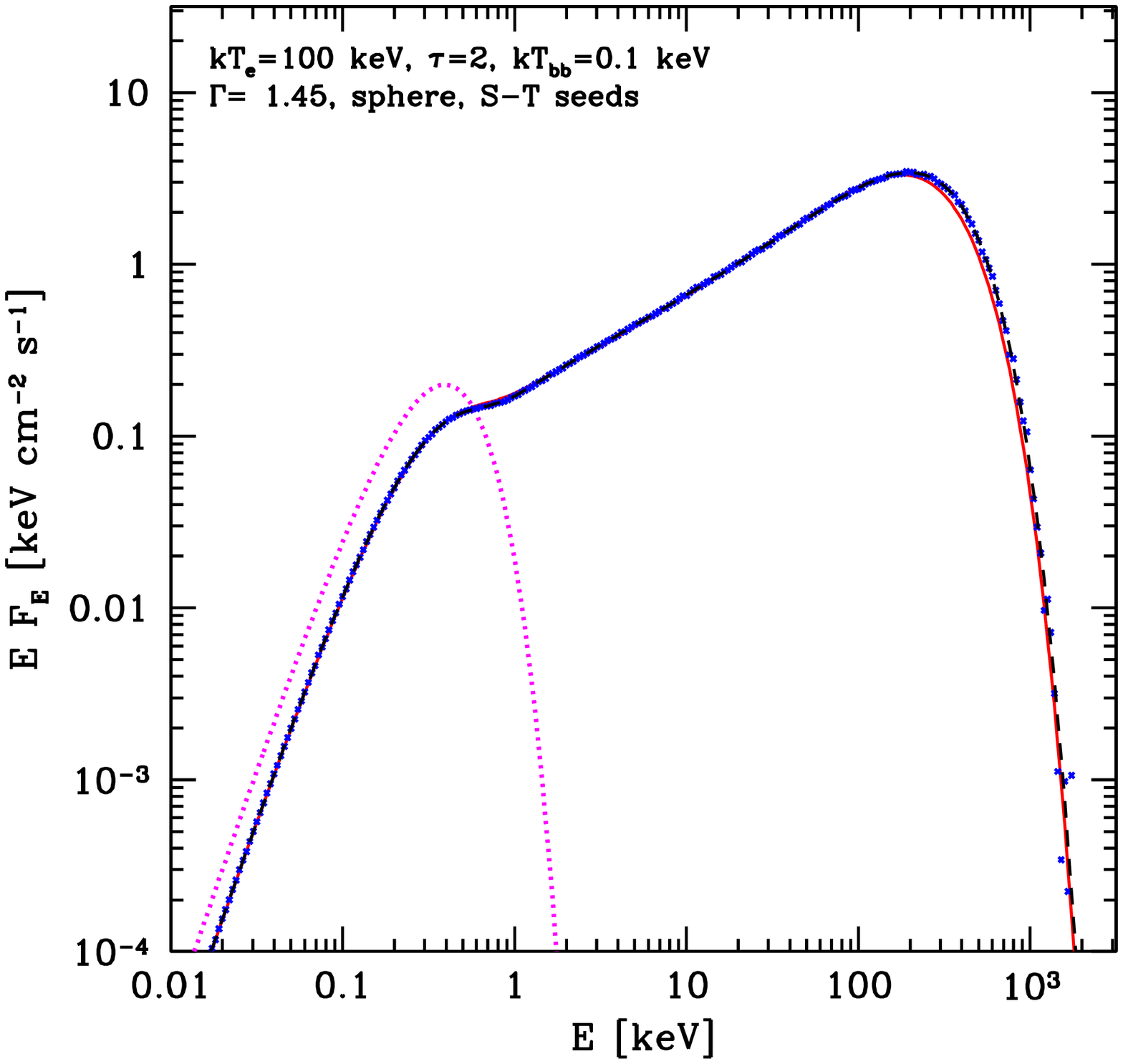}}
\centerline{\includegraphics[width=7.7cm]{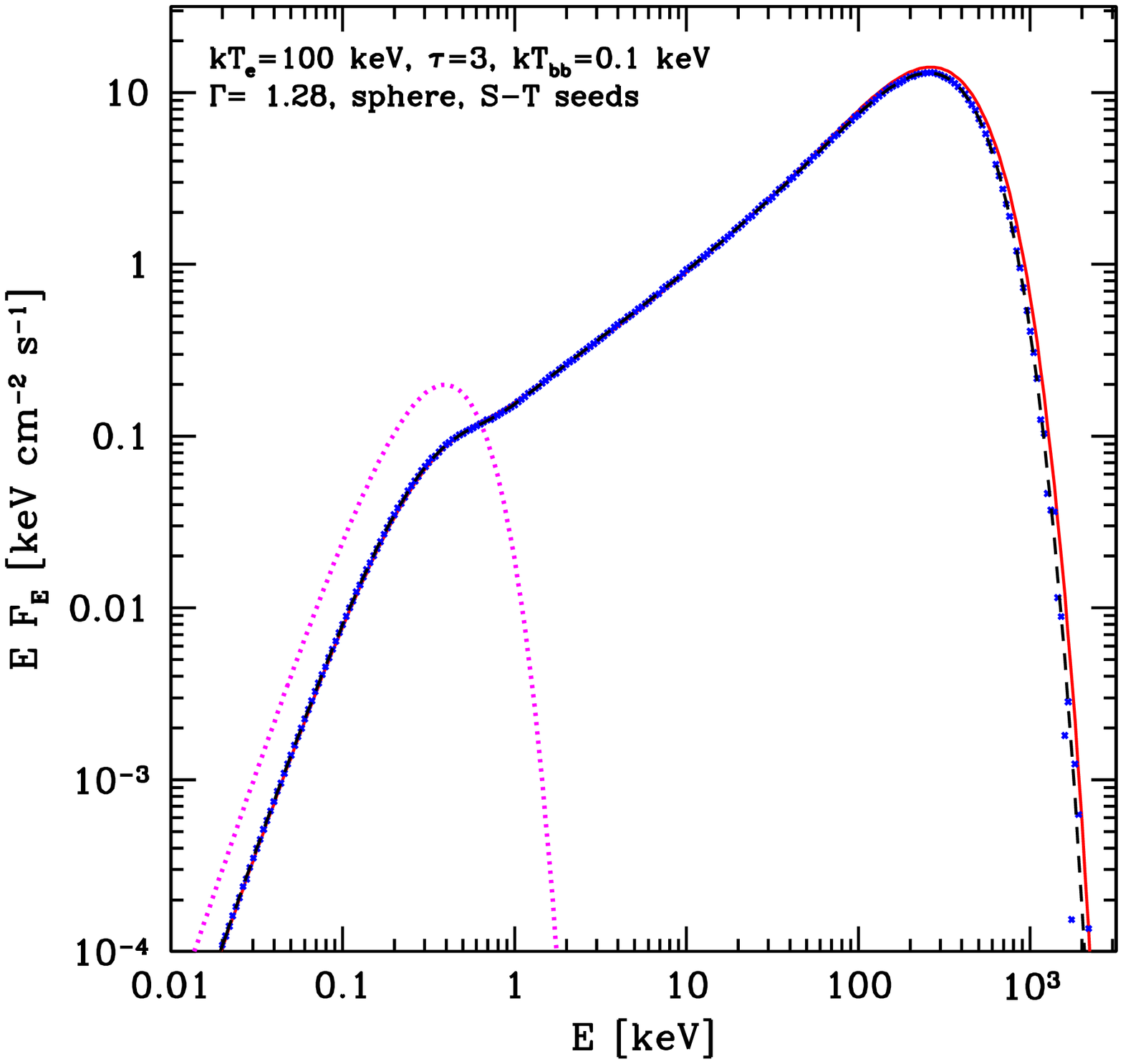}
\includegraphics[width=7.7cm]{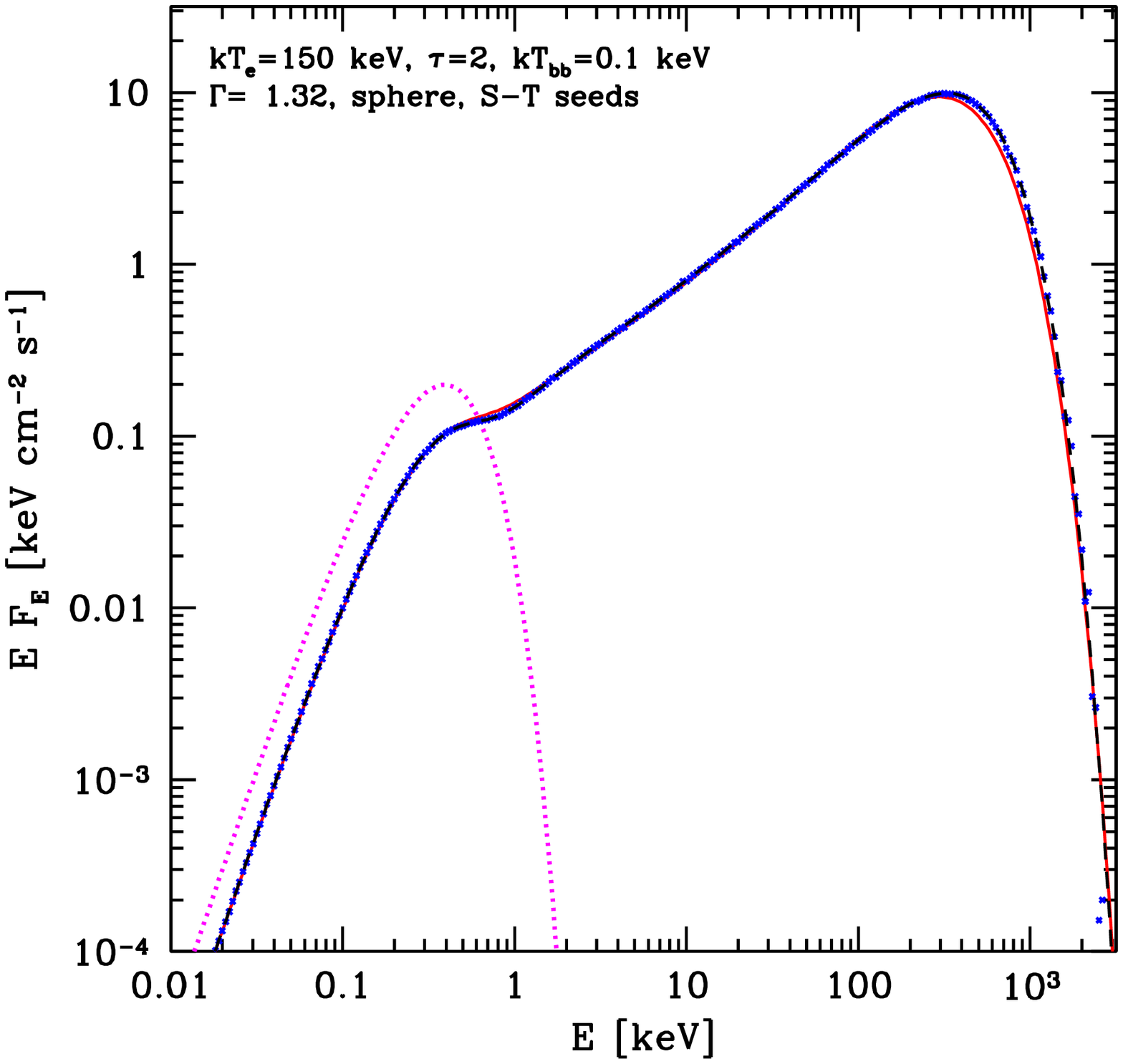}} 
\caption{Continued. 
}
\end{figure*}

\begin{figure*}
\centerline{
\includegraphics[width=7.9cm]{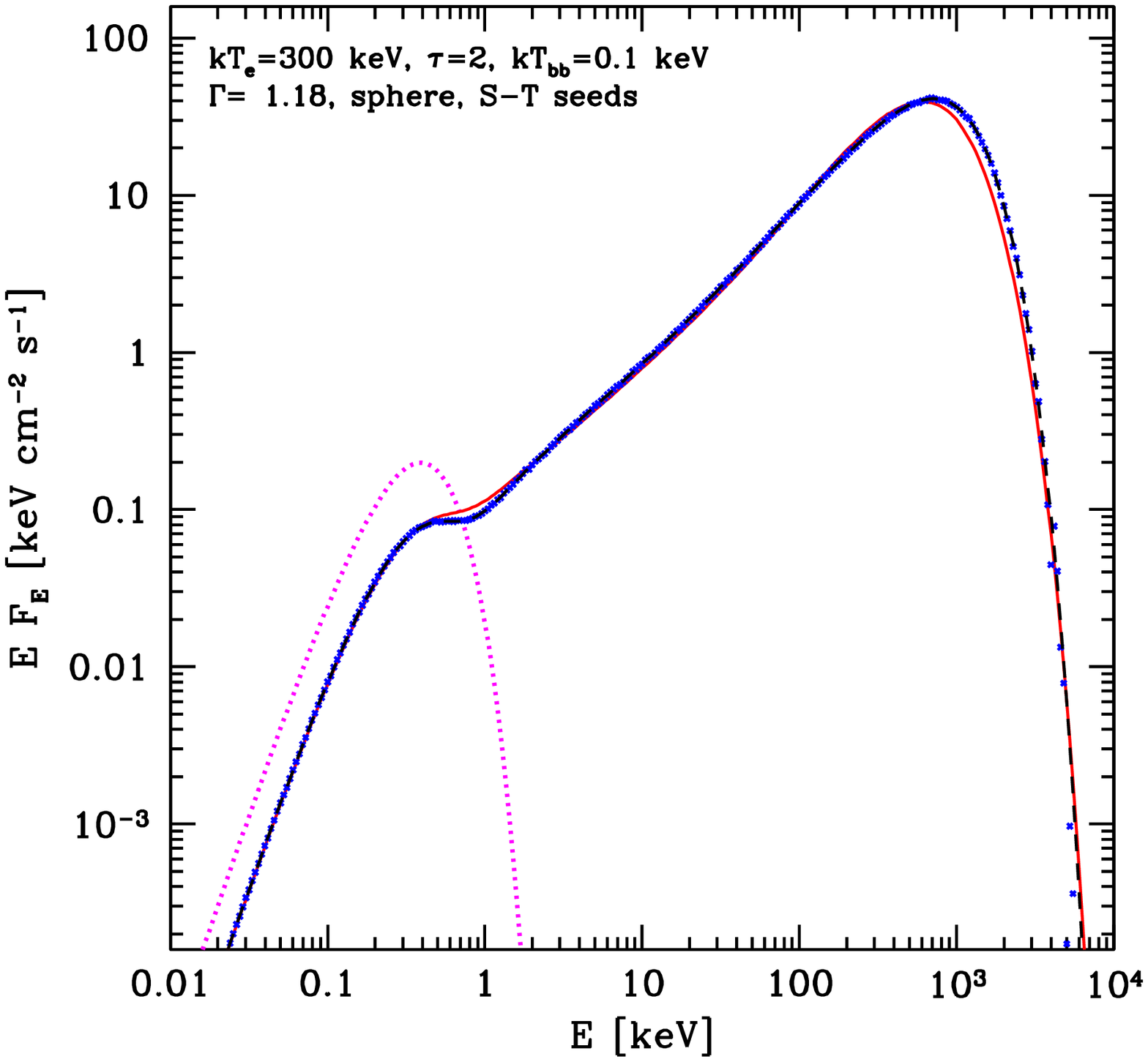}
\includegraphics[width=7.9cm]{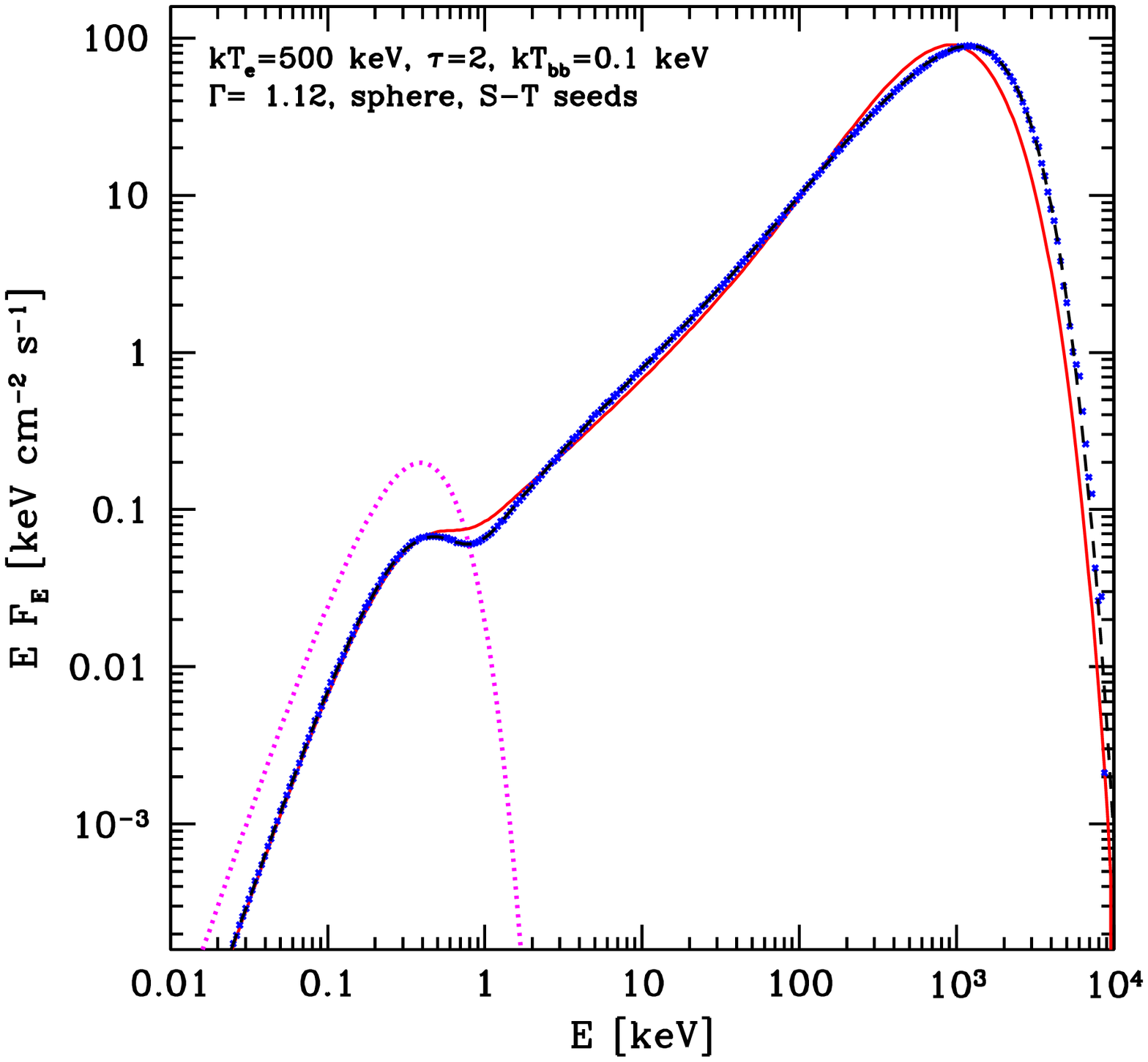}}
\centerline{
\includegraphics[width=7.9cm]{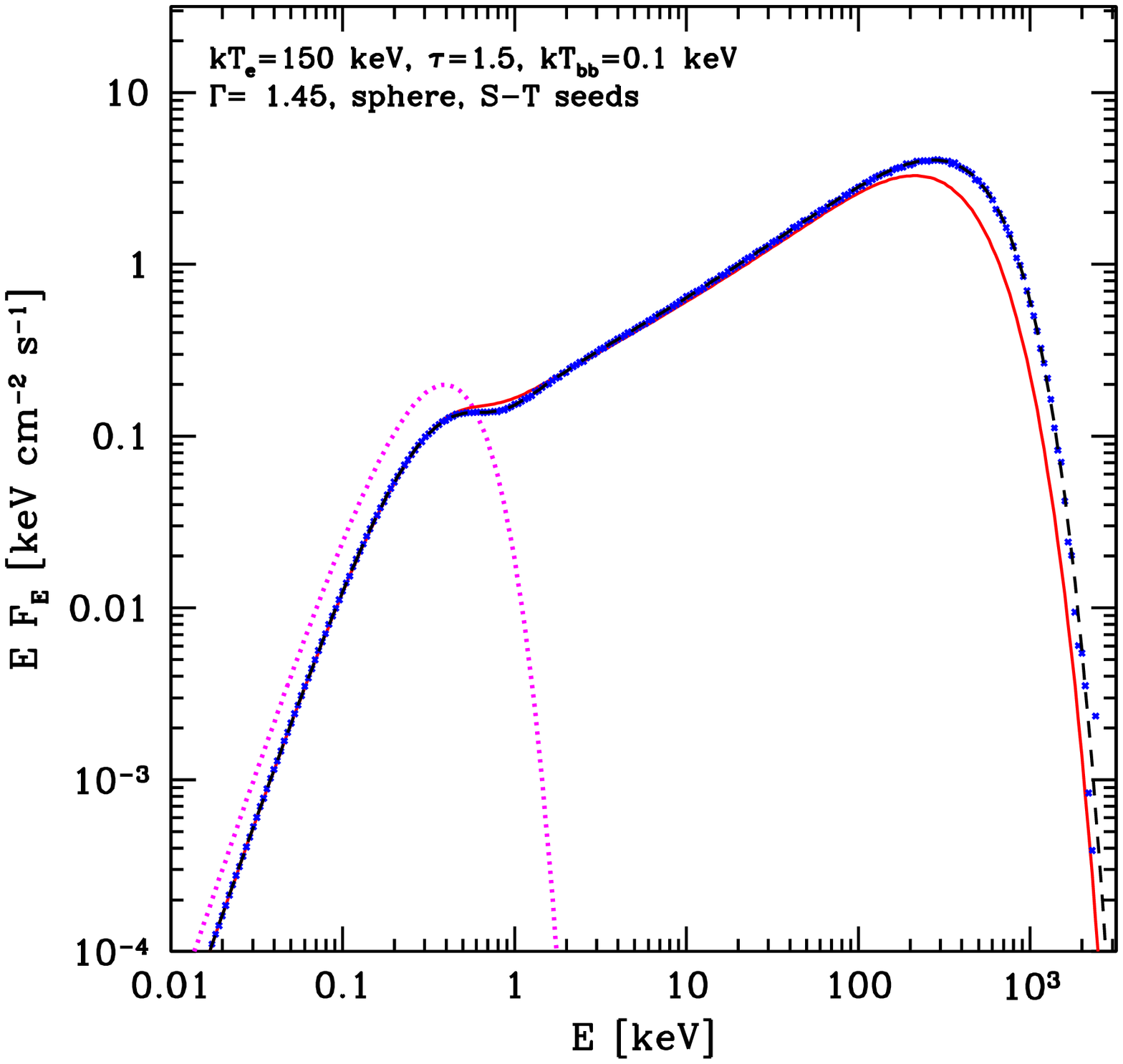}
\includegraphics[width=7.9cm]{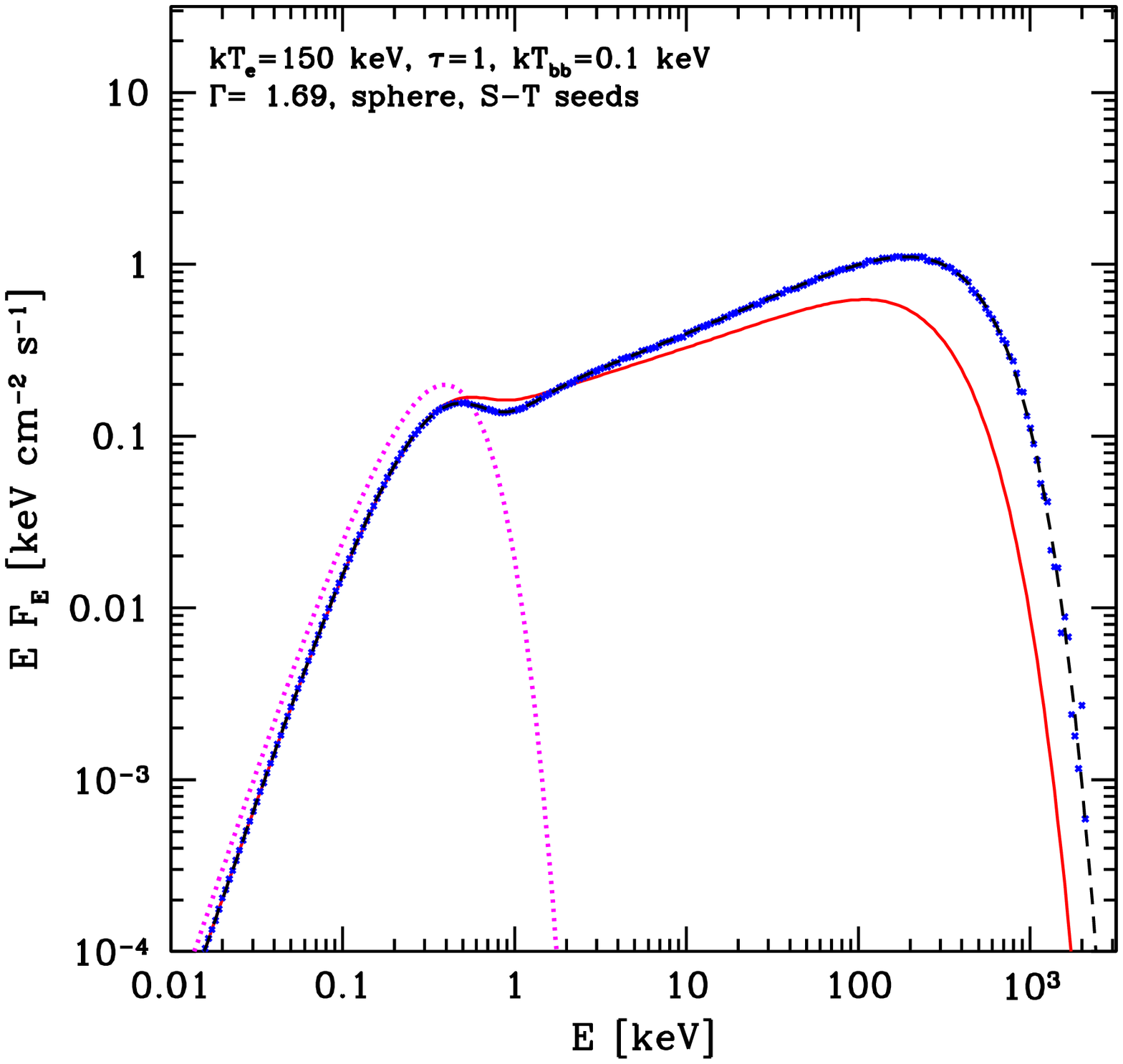}}
\centerline{
\includegraphics[width=7.9cm]{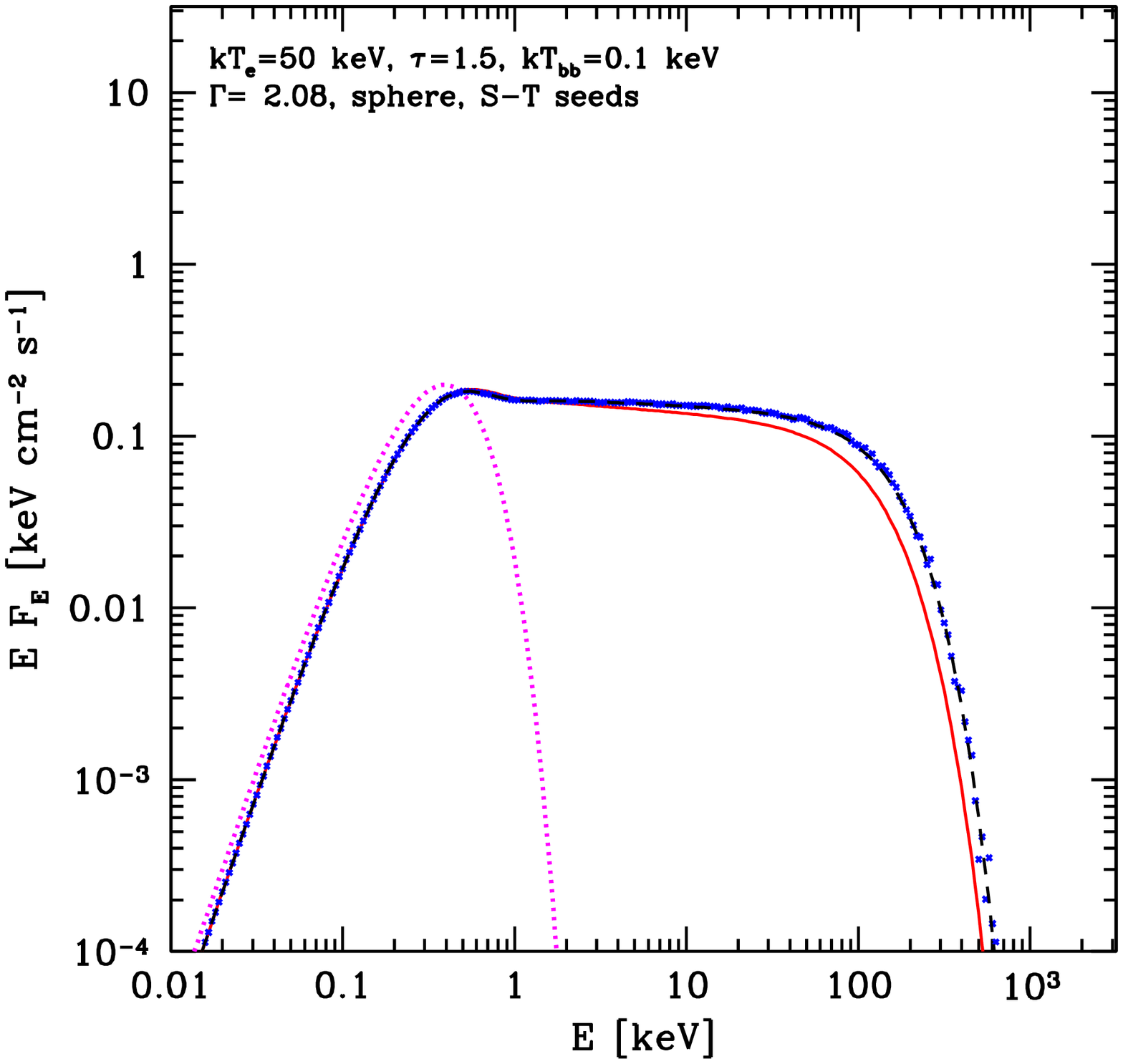}
\includegraphics[width=7.9cm]{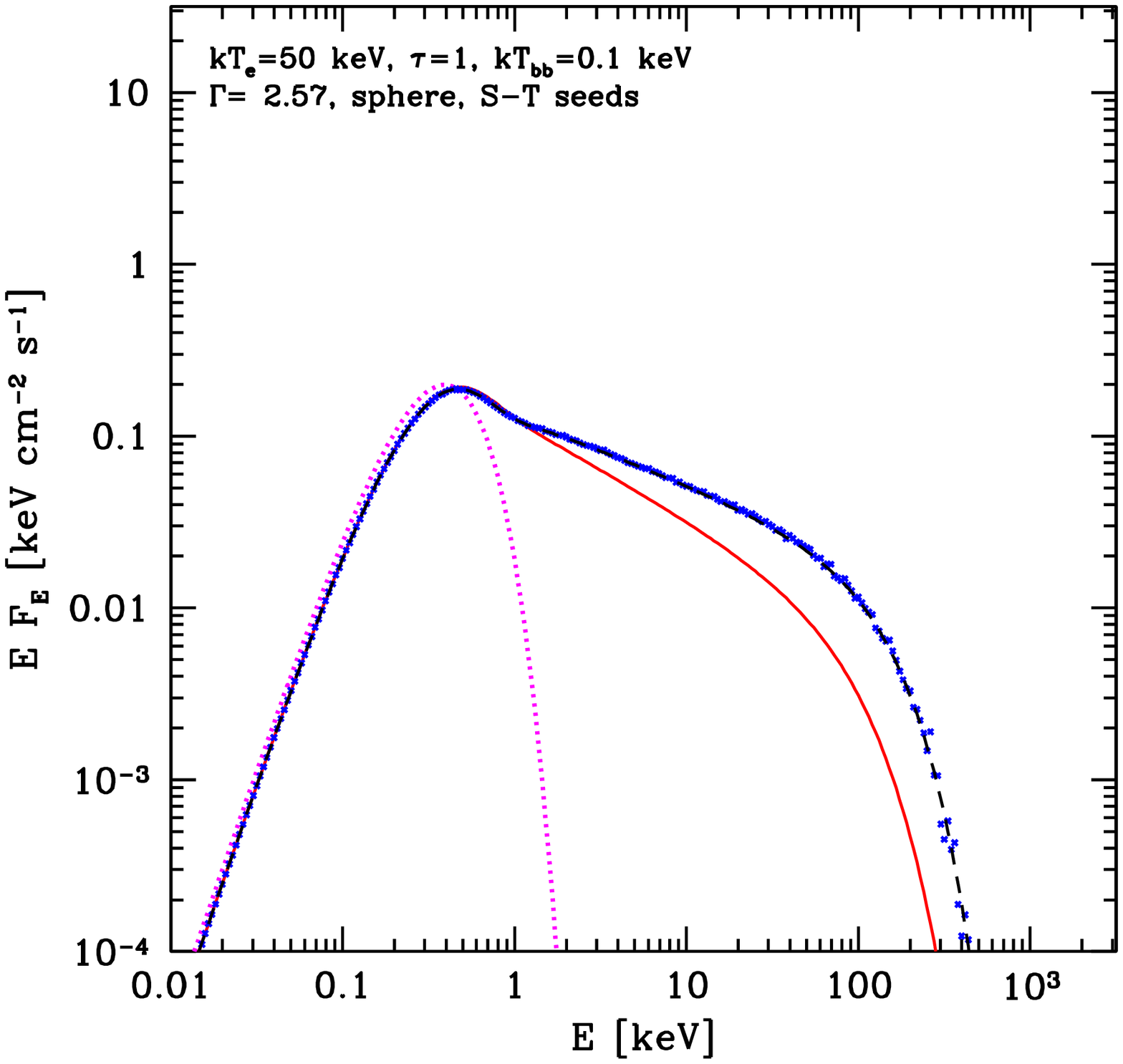}} 
\caption{Comparison of thermal Comptonization spectra calculated using the Monte Carlo method, \compton (blue points), the solution of our kinetic equation, \thcomp (red solid curves), and the ISM, \compps (black dashed curves), for $\tau \leq 2$. The assumptions are the same as in Fig.\,\ref{tau2_5}. We find that \thcomp becomes inaccurate at $\tau =2$ for high values of $kT_{\rm e}\ga 300$\, keV, and for $\tau <2$ at any $kT_{\rm e}$. The ISM results are highly accurate in those regimes.
}\label{tau2_1}
\end{figure*}

When equations (\ref{eq}--\ref{nescape}) are solved for $n(\epsilon)$, the resulting photon rate is given by $\dot n_{\rm esc}$, which assumes that all produced photons undergo Compton scattering. However, since we consider moderate optical depths, we need to take into account that some photons escape unscattered. We thus estimate the escape probability for unscattered photons for our source geometry, and add the corresponding fraction of $\dot n_0$ to $\dot n_{\rm esc}$. We found we can accurately estimate the escaping fraction by taking the geometric average of those for a sphere with homogeneous and central distributions of seed photons. Since Compton scattering conserves the photon number, we can normalize the total observed spectrum to the number of photons equal to that of $\dot n_0$.

We also calculate the power-law slope of the part of the spectrum corresponding to upscattering by the thermal motion dominating over down-scattering. This gives the photon index, $\Gamma$ (where $\dot n_{\rm esc}(\epsilon)\propto \epsilon^{-\Gamma}$) of
\begin{equation}
\Gamma=\sqrt{\frac{9}{4}+\frac{1}{\bar u_{\epsilon=0} \theta\xi(\theta)}}-\frac{1}{2},
\label{Gamma}
\end{equation}
i.e., $\bar u$ is calculated using the Thomson optical depth and $g(\epsilon)=1$. This is analogous to equation (17) of \citetalias{st80}.

\begin{figure}
\centerline{
\includegraphics[height=7.cm]{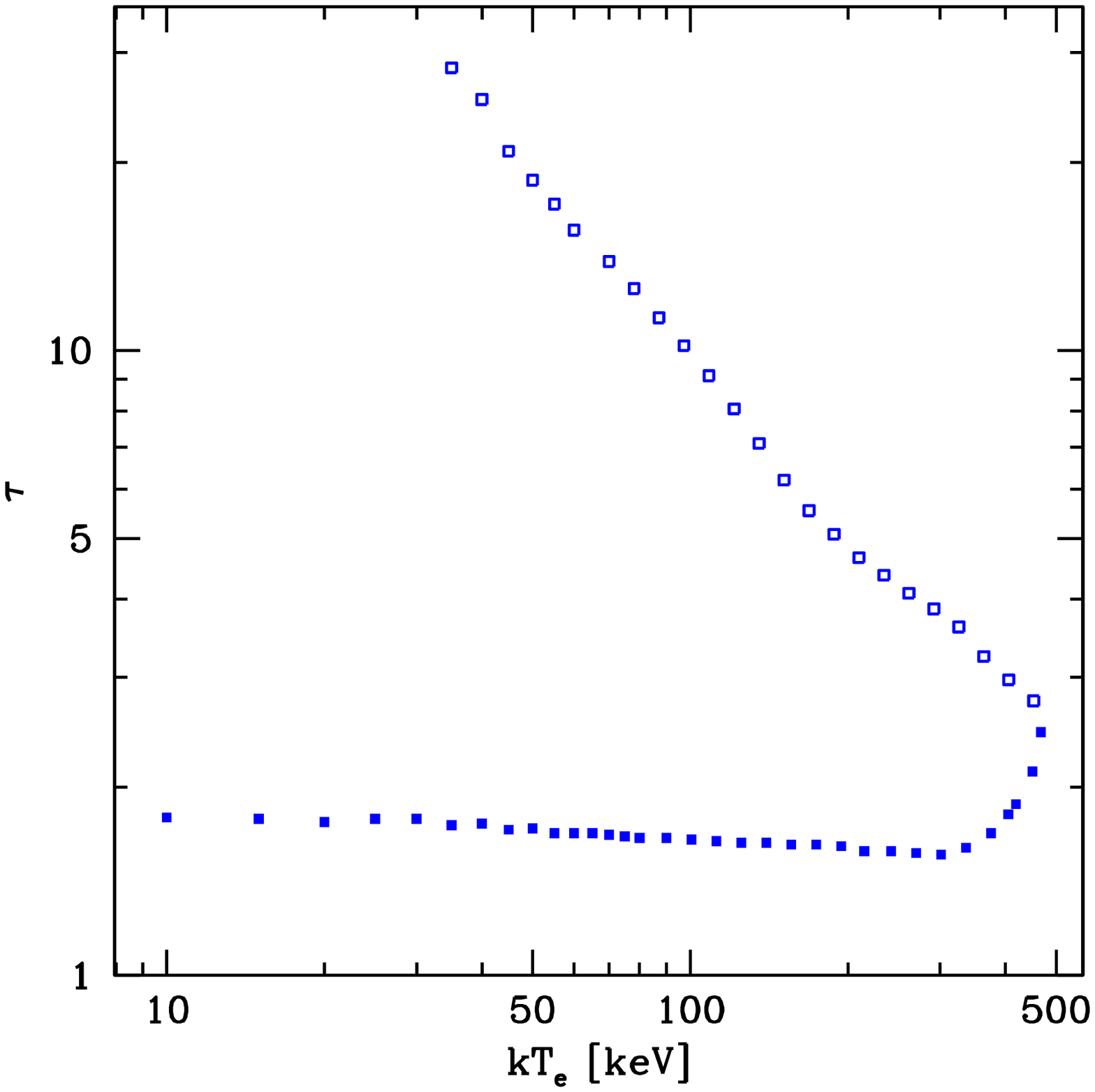}}
\caption{The ranges of $\tau (kT_{\rm e})$ where the accuracy of \thcomp is better than 90 per cent for the case of blackbody seed photons at $kT_{\rm bb}=0.1$\,keV distributed sinusoidally in a spherical cloud with a uniform electron density. The accuracy is measured by the ratio of the \thcomp spectra to those from \compton averaged over the range of $E=(0.1$--$5) kT_{\rm e}$. The lower limits (filled symbols) show $\tau$ below which the plasma becomes too optically thin for the kinetic equation (\ref{eq}--\ref{nescape}) to work, and the upper limits (open symbols) are due to the stationary solution of equation (\ref{nc}) being a Wien spectrum at $kT_{\rm e}\xi(\theta)$ rather than at $kT_{\rm e}$, see Section \ref{thcomp}. 
}\label{thcomp_limits}
\end{figure}

\section{Comparison between different methods}
\label{comparison}

\subsection{Compton upscattering}
\label{up}

Here, we compare the results of \thcomp with those of the fully accurate Monte Carlo method, \compton. We also compare the above results with those of the iterative scattering method (ISM) of \citet{ps96}, as implemented in the public version of the \compps code. This allows us to determine the ranges of validity of \thcomp and \compps. For the comparison, we assume the spherical geometry and a sinusoidal distribution of the seed blackbody photons of \citetalias{st80}, which geometry is assumed in \thcomp (see Section \ref{thcomp}), and which we set for \compton and \compps.

Fig.\,\ref{tau2_5} shows several cases for which \thcomp provides accurate results. Generally, we find it works well for $\tau \geq 1.6$ and $kT_{\rm e}\la 300$\,keV. On the other hand, the ISM method implemented in the \compps code relies on computing the contributions to the spectrum from a limited number of scatterings. Its public version sets their number as $50+4\tau^2$. The accuracy of the code depends then on $\tau$, and it may become inaccurate for large $\tau$ in some cases. Also, since the fractional enhancement of the photon energy in a scattering increases with $kT_{\rm e}$, the Wien peak is reached quicker at a higher $kT_{\rm e}$. When it reached, subsequent scatterings have little effect on the spectrum. Thus, the value of $\tau$ above which \compps becomes inaccurate increases with $kT_{\rm e}$. E.g., for $\tau=3$, the \compps code is much more accurate at $kT_{\rm e}=100$\,keV than at 20 keV, see Fig.\,\ref{tau2_5}. We can see that \compps visibly underestimates the spectrum at the latter $kT_{\rm e}$ even at $\tau=2$. We note, however, that the accuracy of \compps can be increased by increasing the number of iterations, and, in the case of low temperatures, by increasing the grid resolution used in the code.

Then Fig.\,\ref{tau2_1} shows cases where \thcomp starts to provide inaccurate results. At $\tau =2$ and $kT_{\rm e}=300$\,keV, the peak of the spectrum is slightly underestimated, but the disagreement becomes pronounced at $kT_{\rm e}=500$\,keV. Then, \thcomp significantly underestimates the spectrum for $\tau =1$. 

Fig.\,\ref{thcomp_limits} shows the limits of applicability of \thcomp in the $E$-$\tau$ space, using the criterion of fractional accuracy averaged over the $E=(0.1$--$5) kT_{\rm e}$ range. The shown lower limits, corresponding to the 90 per cent average accuracy, demonstrate that \thcomp is applicable at $\tau \ga 1.6$ up to $kT_{\rm e}\la 300$ keV. At $\tau=2$, the accuracy is achieved up to $kT_{\rm e}\approx 450$ keV. On the other hand, the upper limit are due to equation (\ref{nc}) having the Wien stationary solution at $kT_{\rm e}\xi(kT_{\rm e})$ rather than at $kT_{\rm e}$, as discussed in Section \ref{thcomp}. However, very hard spectra showing pronounced Wien peaks have not been observed from accreting X-ray sources.

We stress that in our search for an efficient relativistic/low $\tau$ kinetic equation we have tested a large number of different cases. In particular, we have tested placing the correction coefficient $\xi$ in front of the square brackets in equation (\ref{nc}), which would solve the problem of the stationary solution, $\dot n_{\rm C}\equiv 0$, of that equation being different from Wien at $T_{\rm e}$. However, this has led to underestimates of the high-energy cutoff at mildly relativistic temperatures and moderate optical depths. As the result of our extensive search, we have found that the set of equations given in Section \ref{thcomp} provides the best overall results, including those for down-scattering, see Section \ref{down} below. 

\begin{figure}
\centerline{
\includegraphics[width=\columnwidth]{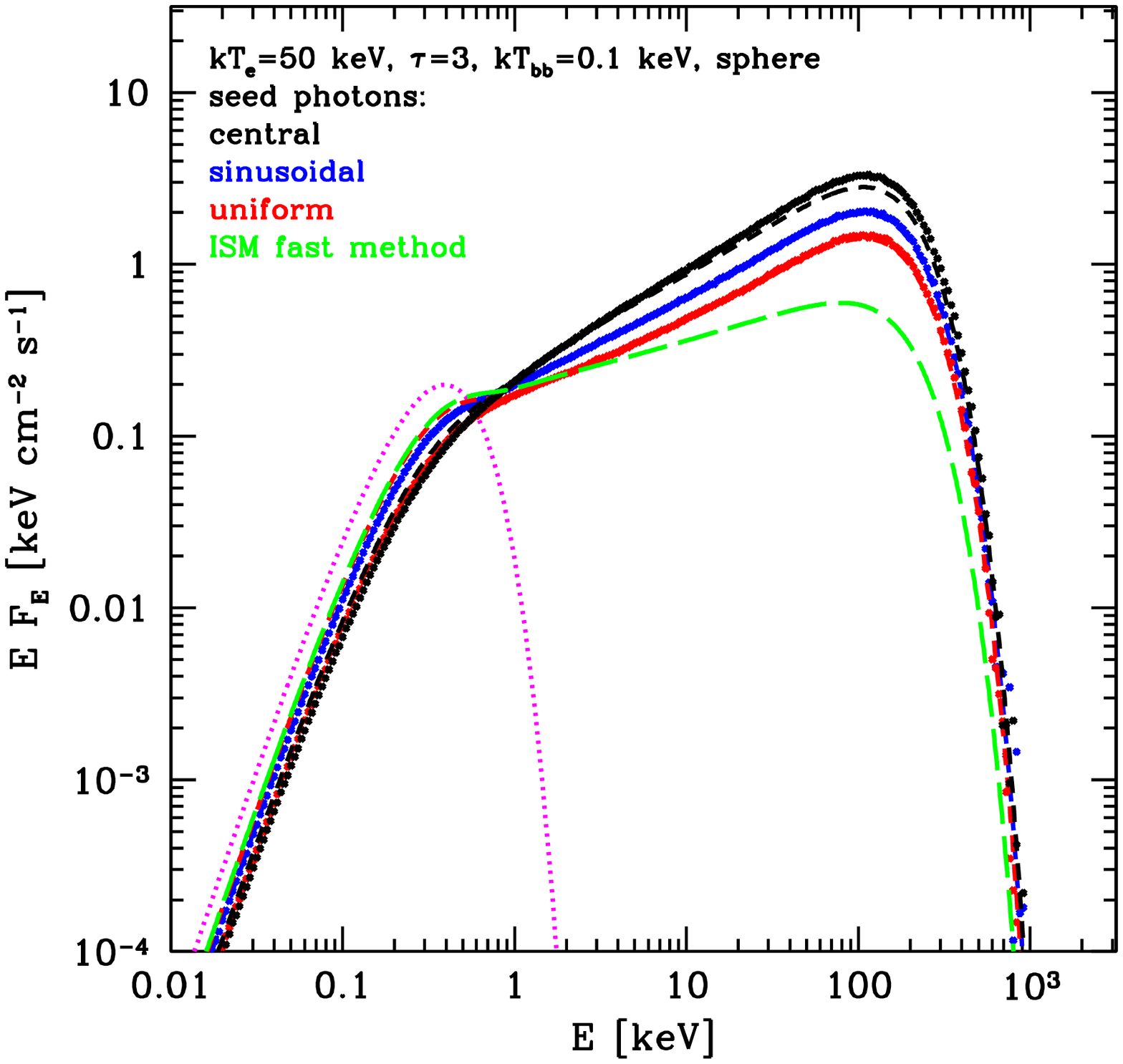}}
\caption{An example of the effect of the spatial distribution of the seed photons on the thermal Comtonization spectra in spherical geometry. Their energy distribution is given by a blackbody at $kT_{\rm bb}=0.1$ keV, shown by the magenta dots. The points are from the Monte Carlo method, \compton, and the dashed curves are from the ISM method, \compps. The shown cases are the central spatial distribution (the hardest spectrum, black), uniform (the softest spectrum, red) and according to the sinusoidal law of \citetalias{st80} (middle, blue). The Monte Carlo points form almost continuous curves in all cases, and the middle and lowest dashed curves almost completely overlap with them. Only for the central seed photons \compps slightly underestimates the Monte Carlo spectrum. Then, the green long-dashed line shows the spectrum assuming the fast ISM method for spherical geometry (\texttt{geom=0} in \compps), which is softer than any of the actual spectra.
}\label{seed_geo}
\end{figure}

We then show an example of the effect of the spatial distribution of the seed photons on the emitted spectra, see Fig.\,\ref{seed_geo}. We see that changing that distribution from central to uniform leads to relatively minor changes of the X-ray slope. The changing spectra can be approximately reproduced by changing the optical depth in a given seed-photon distribution. This follows from the spectral form at a given temperature being governed by the distribution of the number of scatterings, and, approximately, by their average number, see \citet{hua95}. For example, the spectrum at $\tau =3$ for the central distribution can be approximately reproduced by the case with the sinusoidal distribution at $\tau =3.4$. Given that the actual geometry of astrophysical sources is not well determined, the actual value of $\tau$ is dependent on the geometry, while the values of $kT_{\rm e}$ and $\Gamma$ can be determined for a given spectrum relatively uniquely. We also show the spectrum for the fast ISM method, turned on by setting \texttt{geom=0} in \compps (green long-dashed curve). We see that it is softer than all three actual spectra. However, it still reproduces an actual Comptonization spectrum, albeit for a lower $\tau$. In the present case, it correspond to a Comptonization spectrum with $\tau\approx 2.5$. 

\begin{figure}
\centerline{
\includegraphics[width=7.6cm]{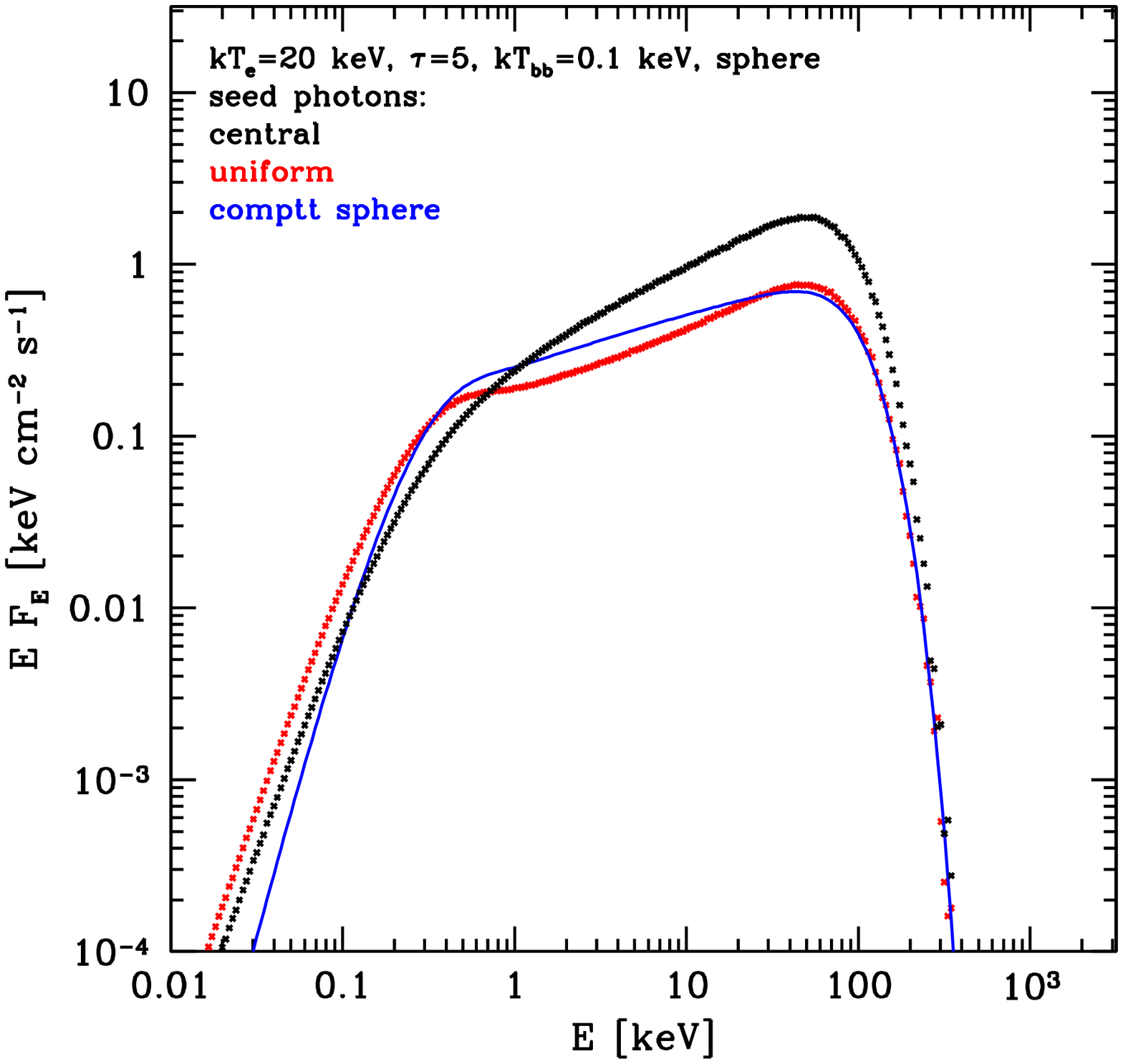}}
\centerline{
\includegraphics[width=7.6cm]{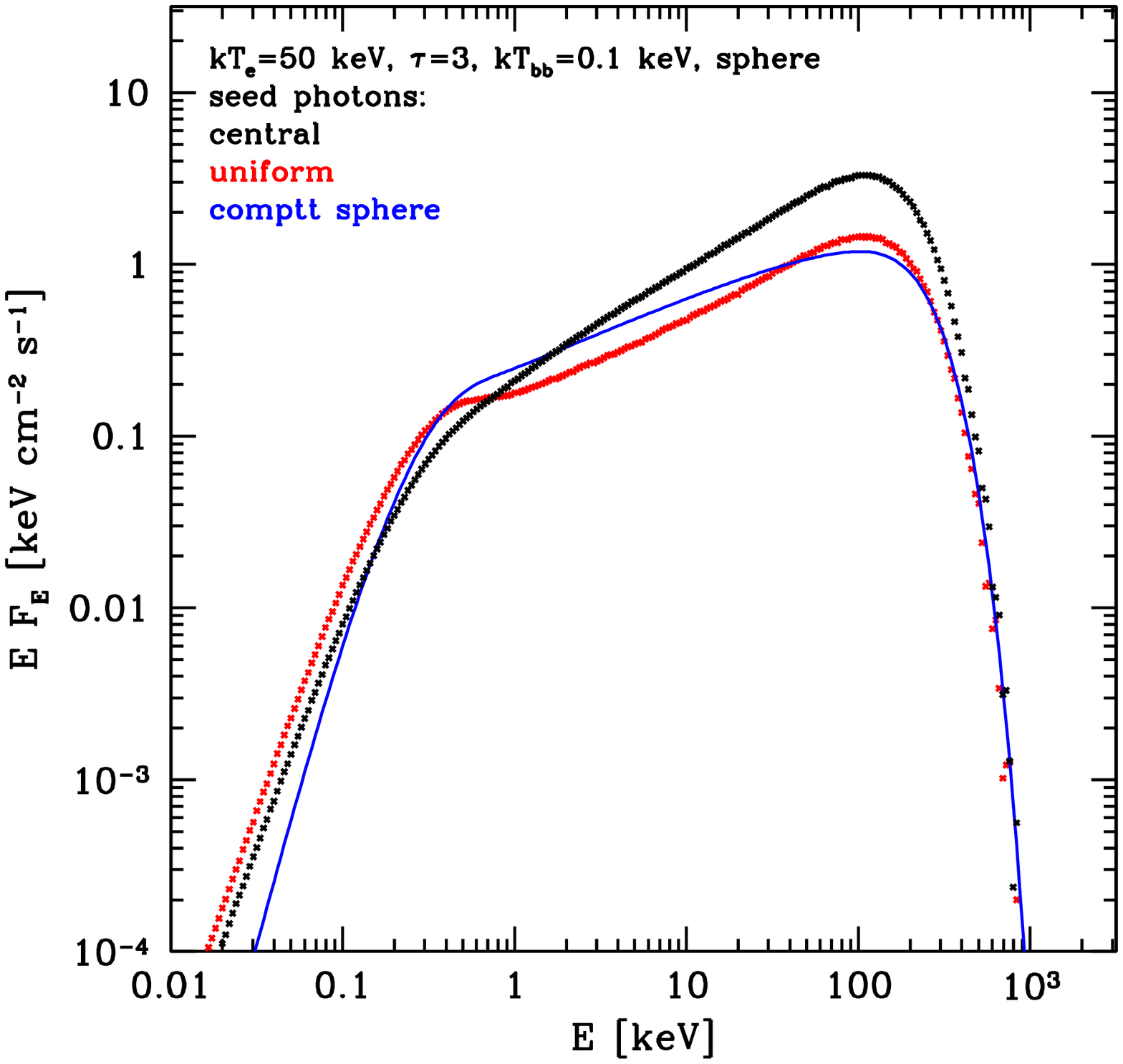}}
\centerline{
\includegraphics[width=7.6cm]{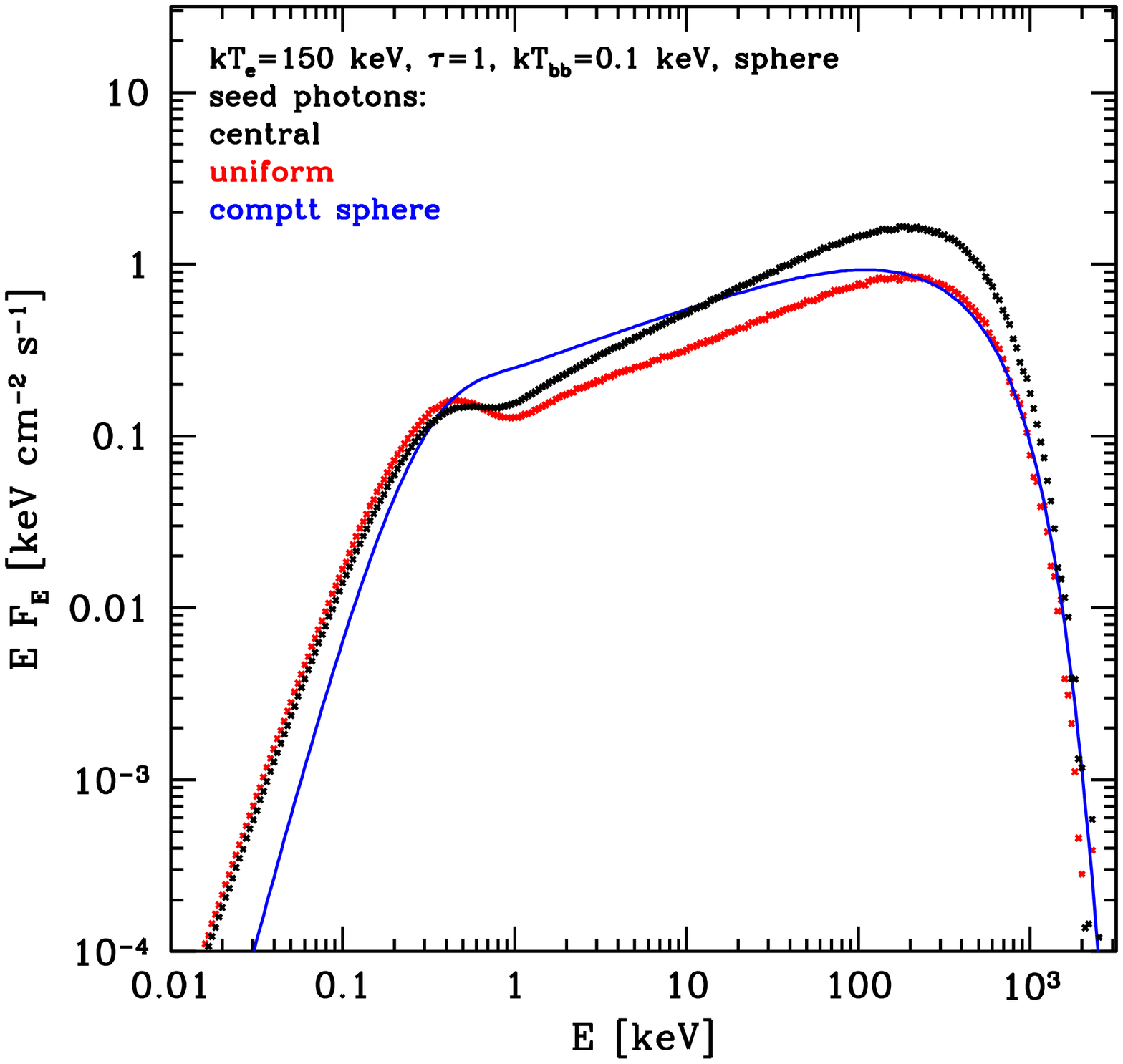}}
\caption{Comparison of some example spectra from the \comptt model (blue solid curves) with the corresponding Monte Carlo spectra in the spherical geometry with the central (upper black dots) and uniform (lower red dots) distributions of seed photons. We see that the \comptt spectra are in all three cases significantly softer than the actual spectra for either distribution of seed photons.
}\label{comptt}
\end{figure}

We also compare our Monte Carlo results in the spherical geometry with those of the \comptt model \citep{titarchuk94,titarchuk95,hua95}. Fig.\,\ref{comptt} shows that the spectra from \comptt for $kT_{\rm e}$ from 20 to 150\,keV are significantly softer than the actual spectra for either distribution of the soft seed photons, though they are generally closer to the case of the uniform distribution. We have found that the \comptt model is less accurate than \thcomp in its range of validity, i.e., for $\tau\ga 1.6$. The relative inaccuracy of \comptt is also illustrated in figs.\ 3--6 of \citet{hua95} (as we noticed in Section \ref{intro}).

\subsection{Compton down-scattering}
\label{down}

\begin{figure}
\centerline{
\includegraphics[width=7.6cm]{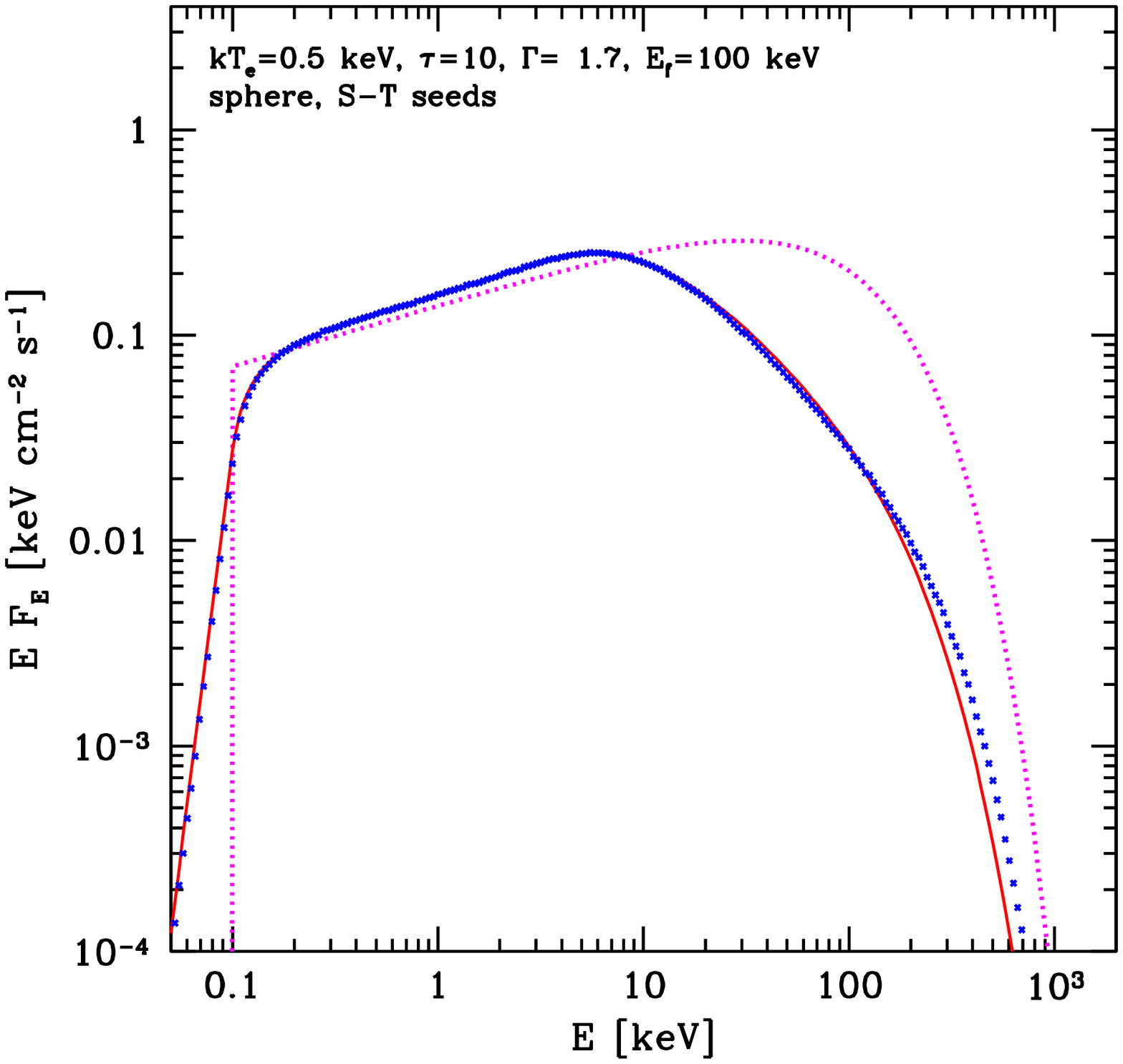}}
\centerline{
\includegraphics[width=7.6cm]{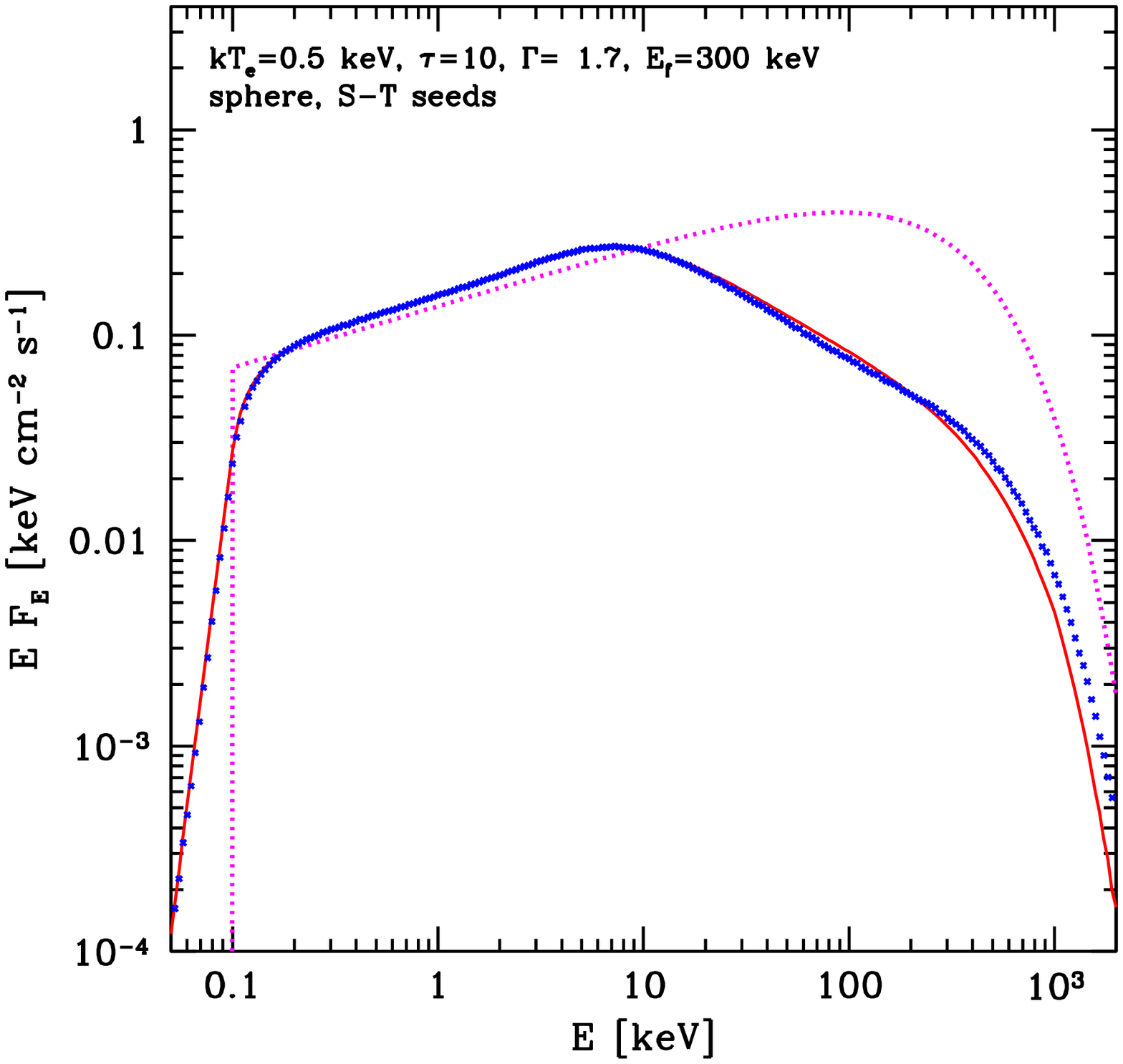}}
\centerline{
\includegraphics[width=7.6cm]{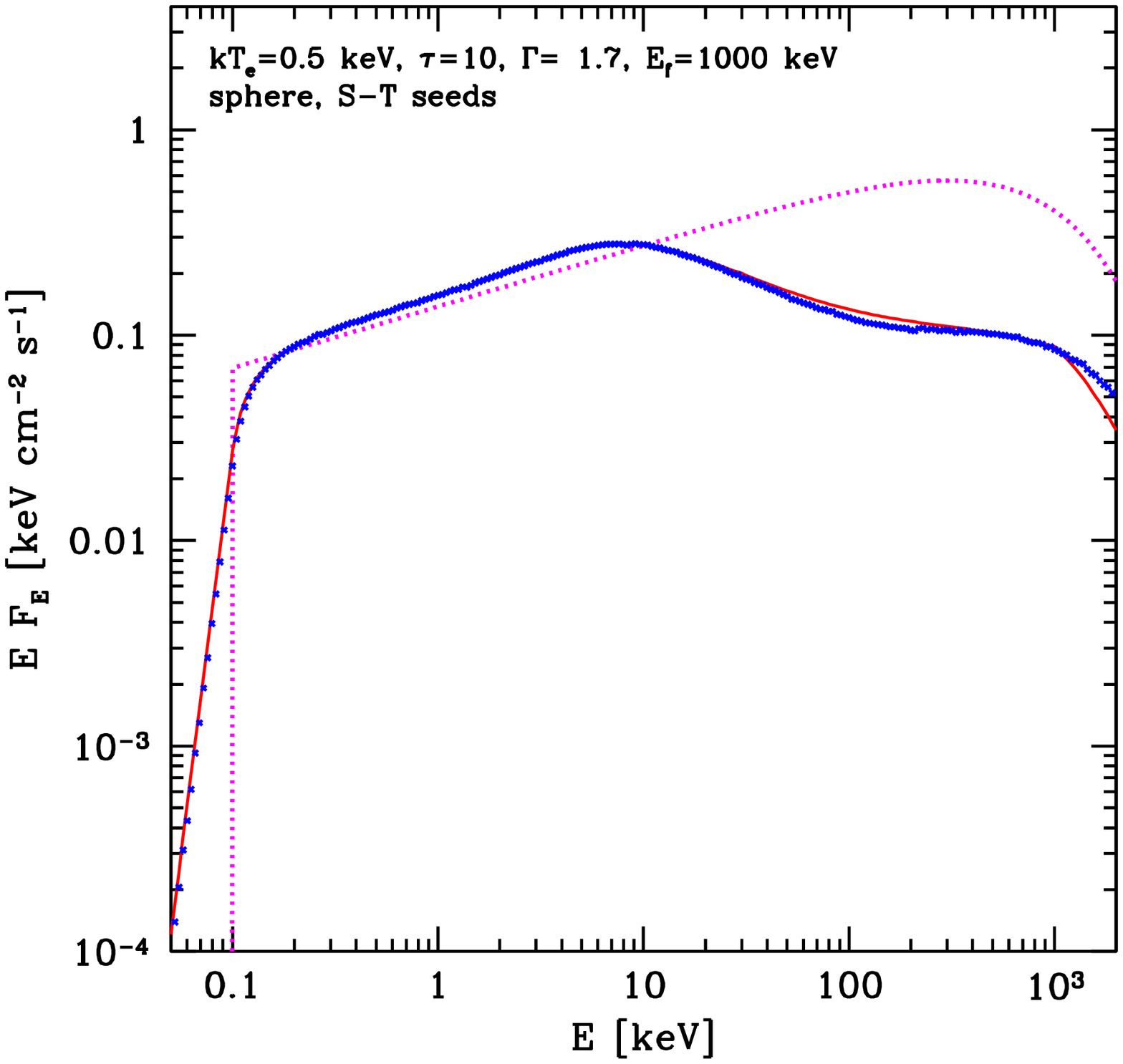}}
\caption{Thermal Comptonization of hard e-folded power-law seed spectra with $\Gamma=1.7$ and the minimum energy of 0.1\,keV (shown by magenta dotted curves) by cold electrons, $kT_{\rm e}=0.5$\,keV, with $\tau =10$. We compare the Monte Carlo method, \compton (blue points), with the solution of our kinetic equation, \thcomp (red solid curves). The assumptions are the same as in Fig.\,\ref{tau2_5}. We find that \thcomp is highly accurate except for some minor discrepancies in the high-energy tails at energies above a few hundred keV.
}\label{down_cold}
\end{figure}

\begin{figure}
\centerline{
\includegraphics[width=7.8cm]{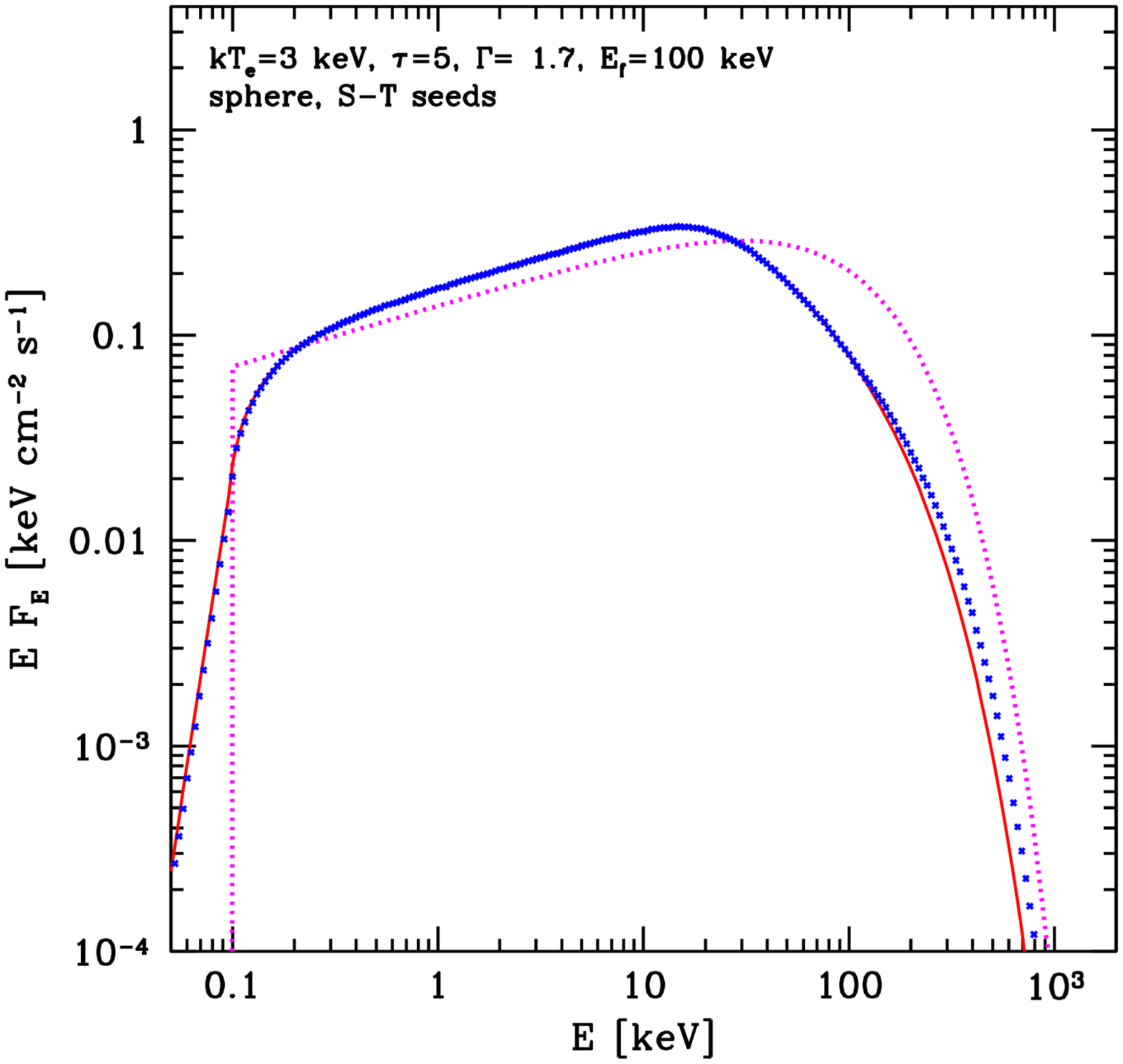}}
\centerline{
\includegraphics[width=7.8cm]{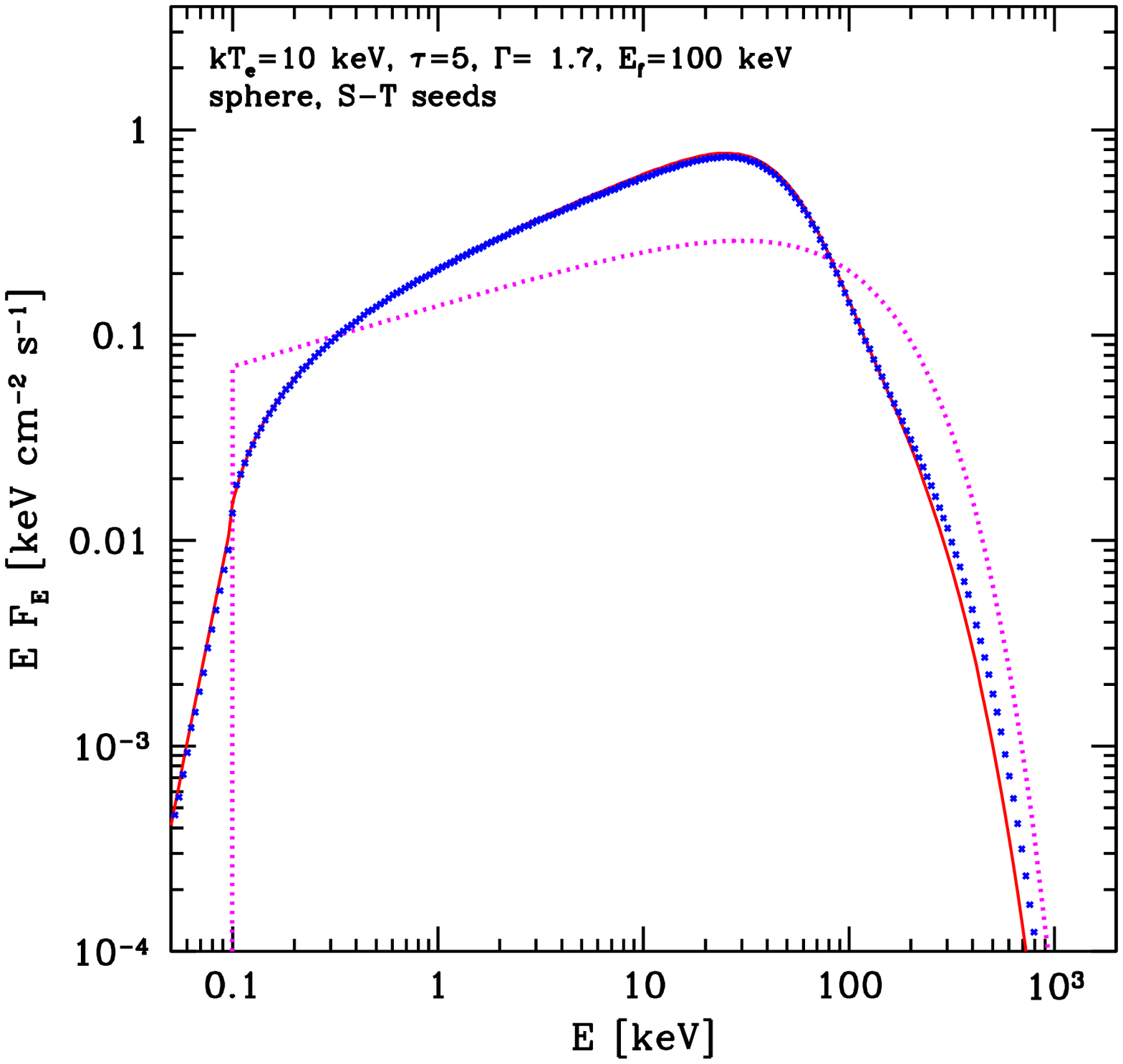}}
\centerline{
\includegraphics[width=7.8cm]{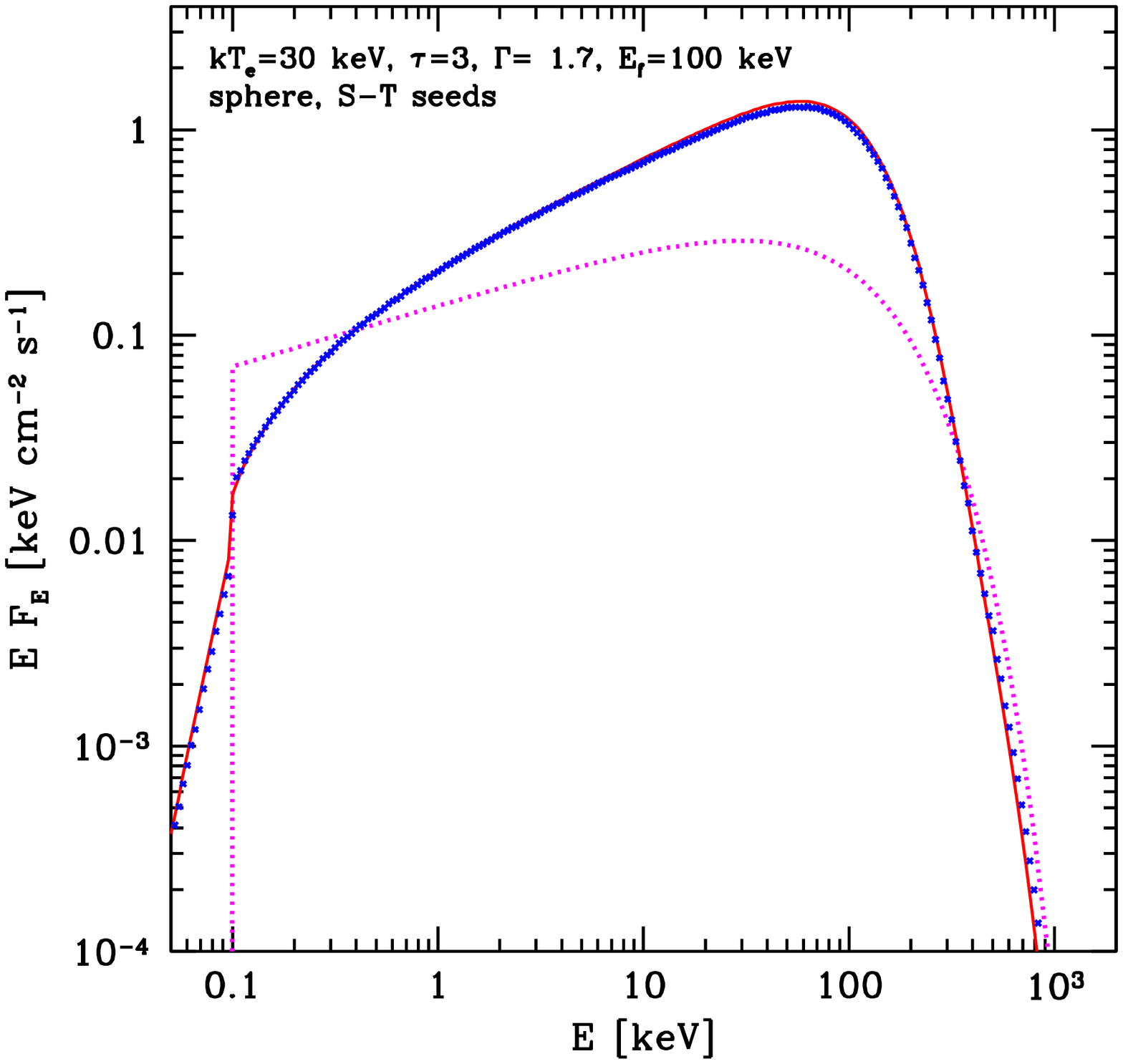}}
\caption{Thermal Comptonization of hard e-folded power-law seed spectra with $\Gamma=1.7$ and $E_{\rm f}=100$\,keV (shown by magenta dotted curves) by warm electrons, $kT_{\rm e}=3$--30\,keV. We compare the Monte Carlo method, \compton (blue points), with the solution of our kinetic equation, \thcomp (red solid curves). The assumptions are the same as in Fig.\,\ref{tau2_5}. We find that \thcomp is highly accurate except for some discrepancies in the high-energy tails.
}\label{down_warm}
\end{figure}

We then consider the applicability of our kinetic-equation results to cases where the characteristic photon energies are much higher than the electron temperature, $kT_{\rm e}$, and the optical depth is high. We compare \thcomp and \compton only, since the ISM method is not applicable in this regime. Fig.\,\ref{down_cold} shows some examples of Comptonization of hard e-folded power-law seed spectra with $\Gamma=1.7$ (note that $\Gamma$ in this Section denotes the index of the e-folded power law) and the e-folding energy of 0.1--1\,MeV passing through relatively cold, optically-thick, electrons. We see an overall excellent agreement between the results of our kinetic equation and the Monte Carlo code.

As it is well known, down-scattering of power-law spectra results in the appearance of a down-scattering break, at which the fractional energy loss per scattering, $\Delta E/E\approx E/m_{\rm e}c^2$, times the average number of scattering, $\bar u\sim \tau ^2$, see equations (\ref{ubar}--\ref{an}), is of the order of unity. In our examples, this break is around 10 keV, which is much lower than the high-energy cutoffs present in the incident spectra. As proposed by \citet*{zmg10}, this effect is likely to explain the low values of the high-energy breaks in the hard state of Cyg X-3. \citet{zmg10} used the Monte Carlo method to calculate model spectra. With our present results, spectra showing such effect can be directly fitted by our \thcomp model as implemented in a convolution version in \xspec.

As mentioned in Section \ref{thcomp}, the equations used do not contain any zero-temperature dispersion term, analogous to that of \citet{ross78}. Still, they give correct results for down-scattering of broad continua. We have repeated the calculations shown in Fig.\,\ref{down_cold} for $kT_{\rm e}$ as low as 50 eV, and found they still reproduce relatively accurately the Monte Carlo results.

Fig.\,\ref{down_warm} then shows some examples with higher values of the electron temperature, $kT_{\rm e}=3$--30\,keV, and lower optical depths, $\tau =3$--5. We still see excellent agreement between the solution of the kinetic equation and Monte Carlo results. We show here only a fraction of the cases we have studied. We find an excellent overall agreement, except at $\tau <2$.

\begin{figure}
\centerline{
\includegraphics[width=8cm]{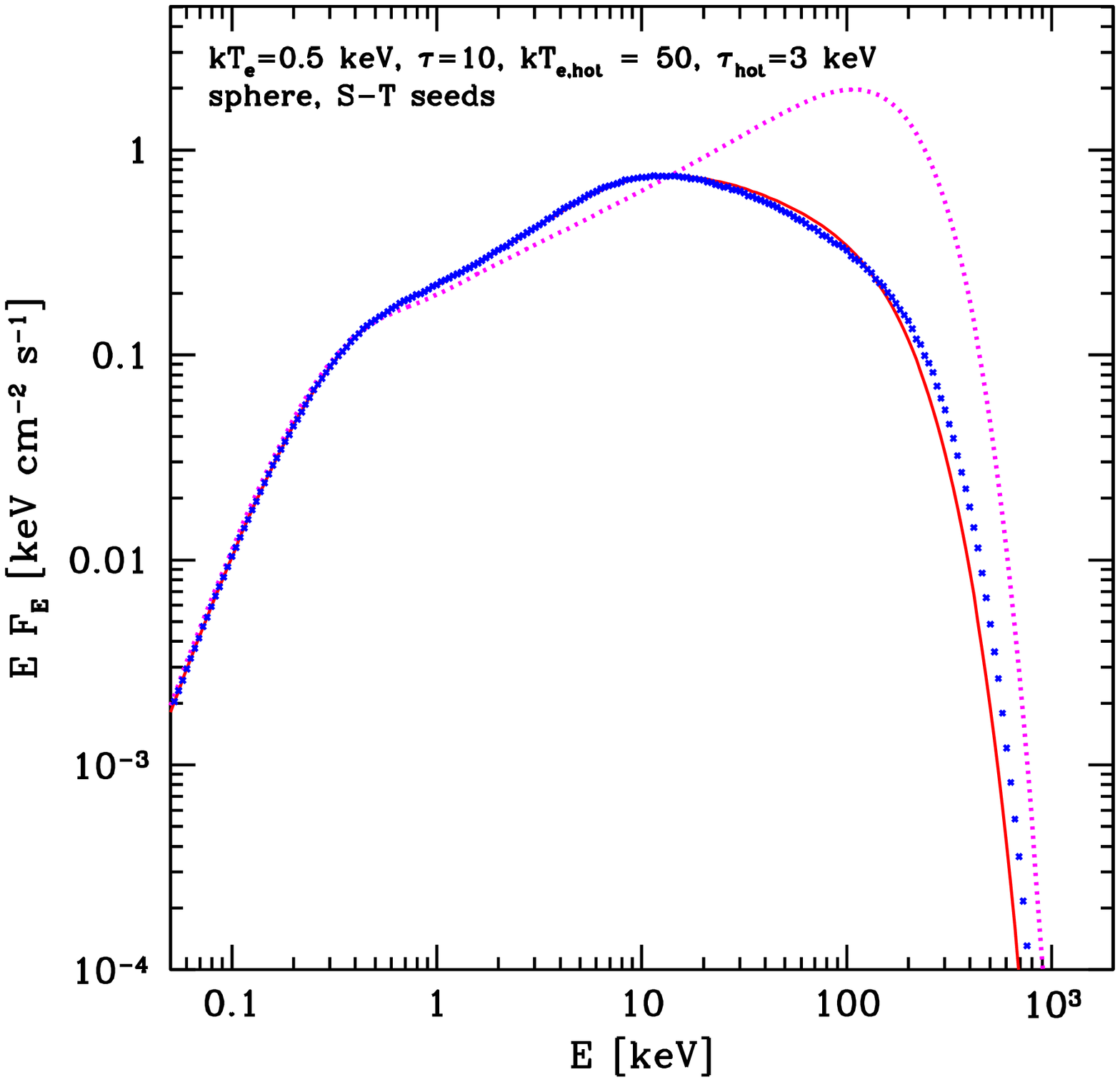}}
\caption{Down-scattering in a cold plasma with $kT_{\rm e}=0.5$\,keV, $\tau =10$, of an initial hard spectrum from thermal Comptonization with $kT_{\rm e, hot}=50$\,keV, $\tau_{\rm hot} =3$, $kT_{\rm bb}=0.1$\,keV (modelled by \compps; magenta dotted curve). The final Comptonized spectrum is modelled by \thcomp (red solid curve) and \compton (blue points). We find that \thcomp gives an overall good approximation, though there are some discrepancies in the high-energy tail.
}\label{down_ISM}
\end{figure}

Fig.\,\ref{down_ISM} then shows the effect of down-scattering in an optically thick cold plasma of a primary spectrum from thermal Comptonization by hot plasma with $kT_{\rm e}=50$\,keV, $\tau =3$, rather than an e-folded power law. We see that our method using the kinetic equation (\thcomp) gives an overall good agreement with the Monte Carlo results.

In Figs.\,\ref{down_cold}--\ref{down_ISM}, we see that \thcomp sligthly underestimates the Monte Carlo spectra at photon energies $E\ga m_{\rm e}c^2$. This is a deficiency of the model, which we have found difficult to remove. Namely, modifications to the coefficients (\ref{cooper}) and (\ref{escape}) that improved that agreement resulted in worsening of the agreement at lower energies, also in the upscattering cases.

\section{The effect of Compton scattering on timing properties}
\label{timing}

\begin{figure}
\centerline{
\includegraphics[height=6.1cm]{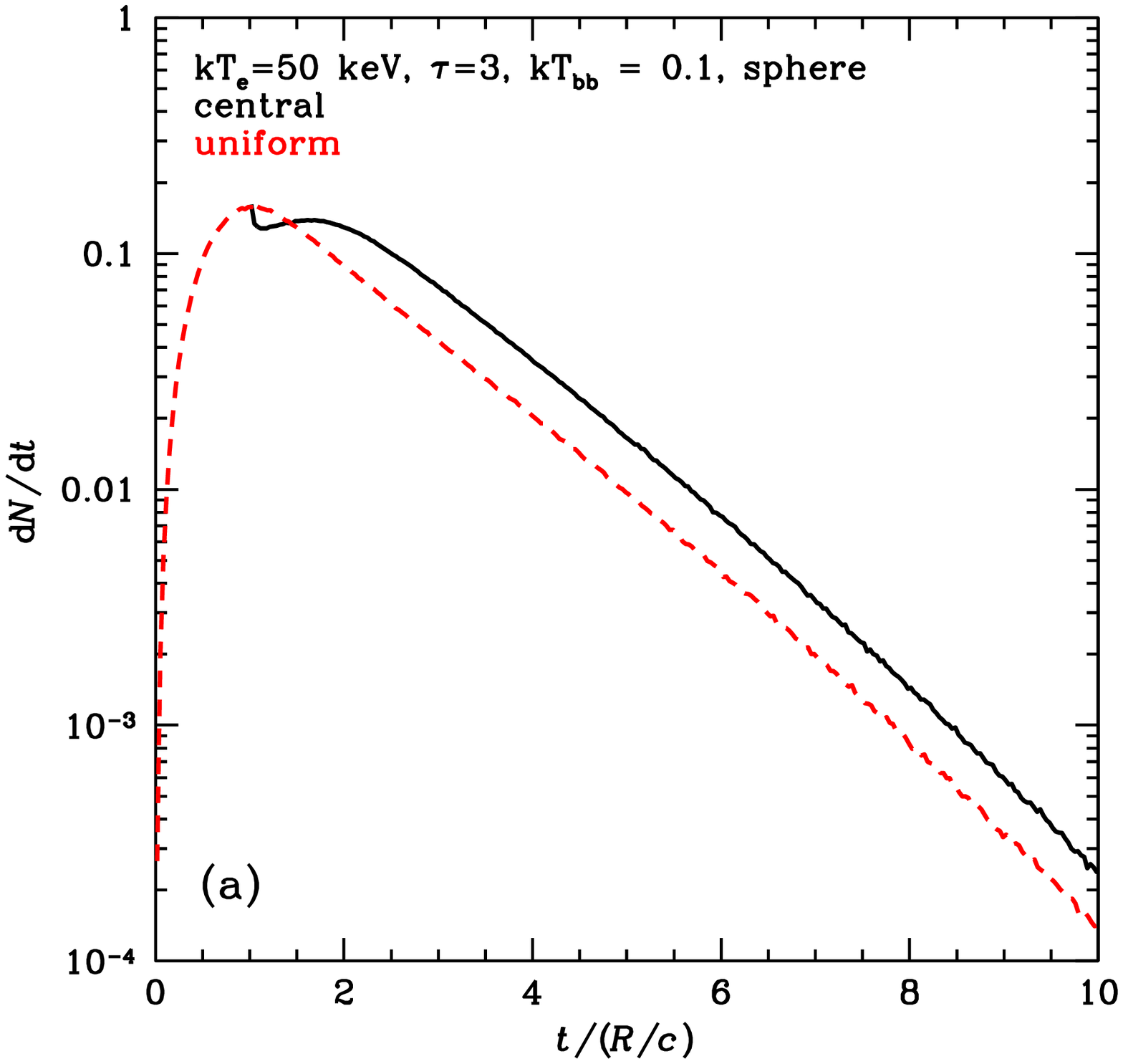}}
\centerline{
\includegraphics[height=6.1cm]{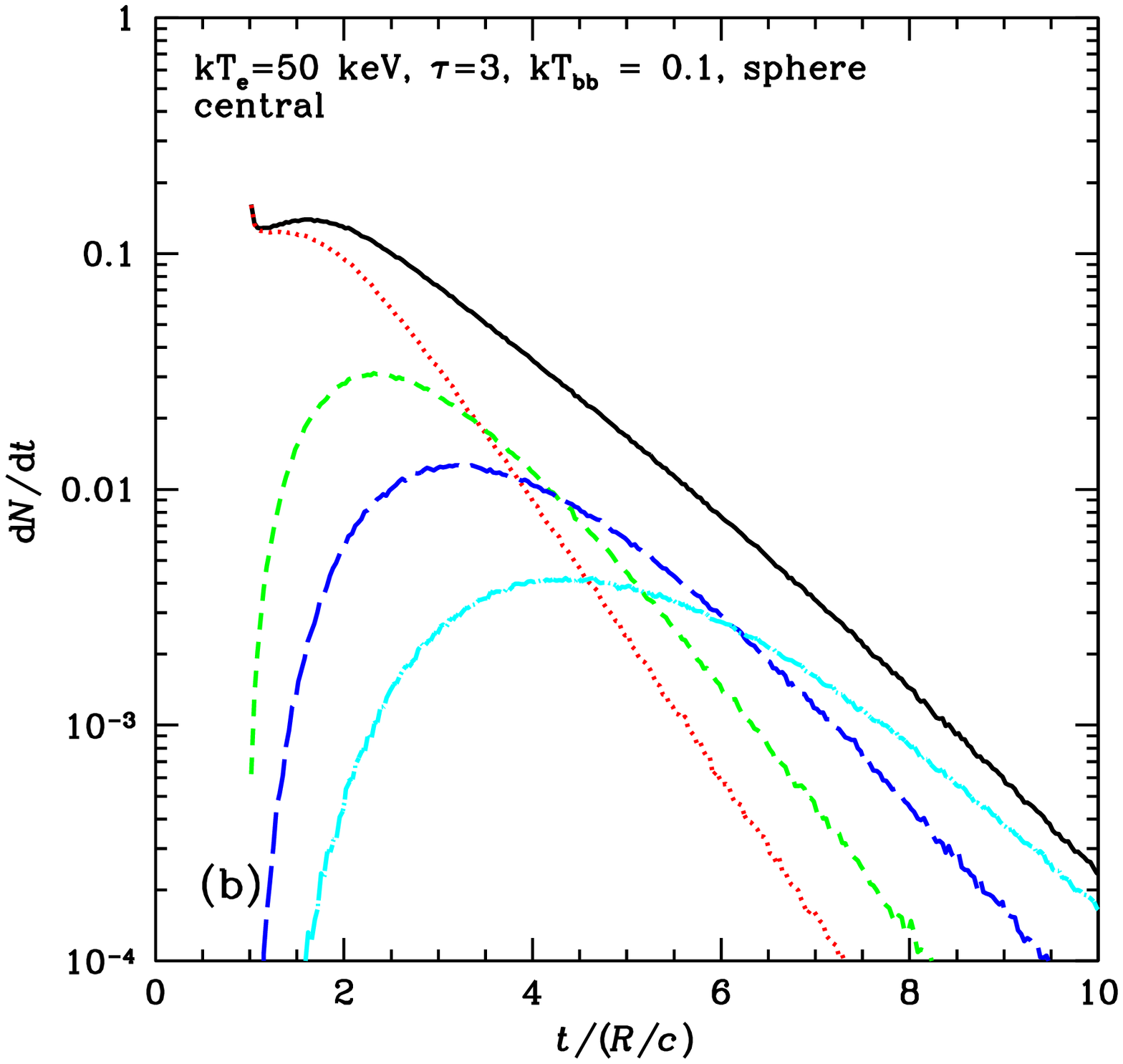}}
\centerline{
\includegraphics[height=6.1cm]{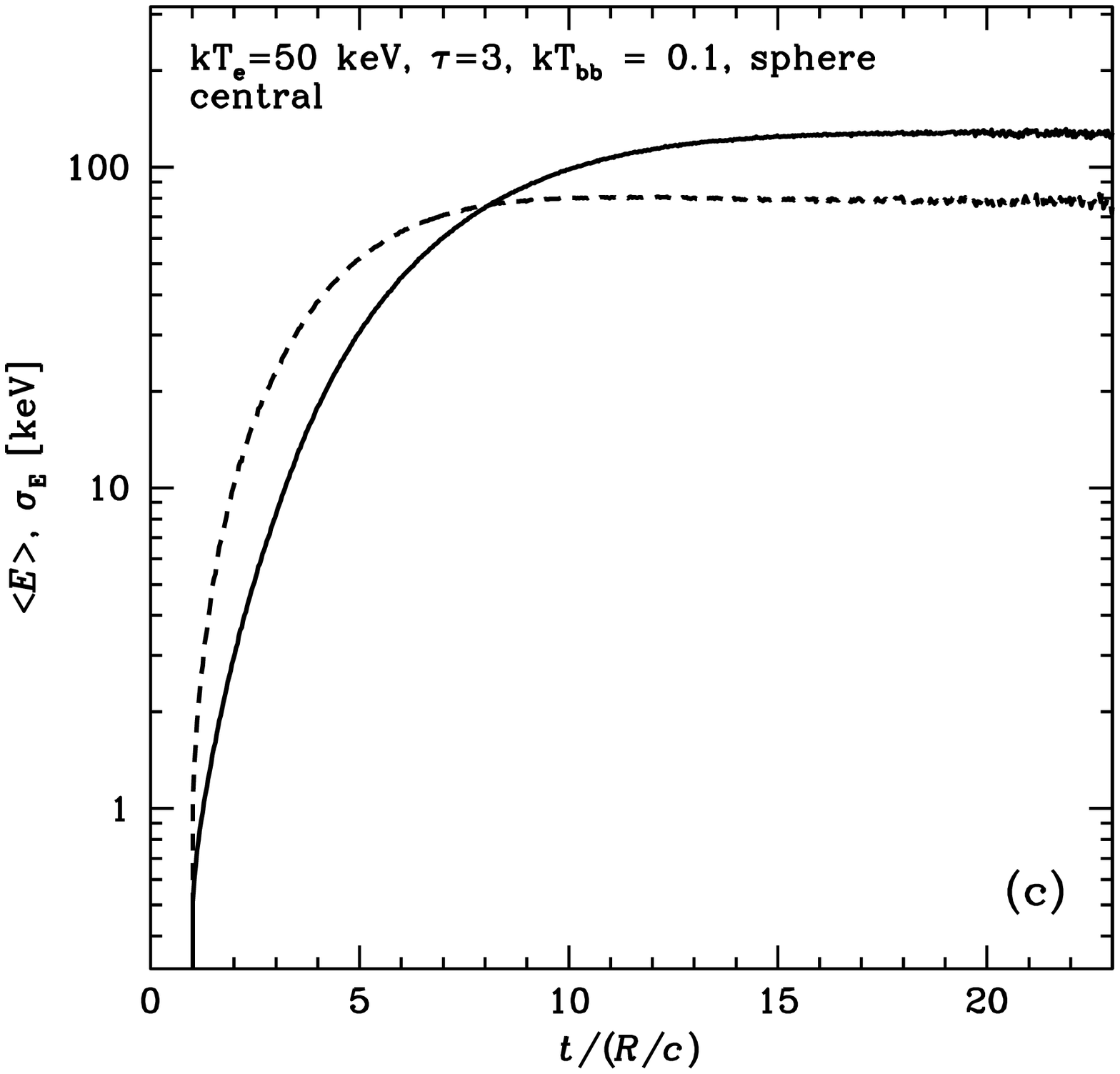}}
\caption{Timing properties of upscattering (modelled by \compton) in a hot plasma with $kT_{\rm e}=50$\,keV, $\tau =3$, of an initial blackbody spectrum with $kT_{\rm bb}=0.1$\,keV. The spectra for this case are shown in Fig.\,\ref{seed_geo}. Only scattered photons are now taken into account. (a) The distributions of the arrival times to a distant observer (Green's functions) for the cases of the seed photons emitted centrally in a sphere (black solid curve) and uniformly distributed (red dashed curve). (b) The distributions of the arrival times for central seed photons divided by the energy range, with the red dots, green short dashes, blue long dashes and cyan dot-dashed curve corresponding to the ranges $<$2, 2--10, 10--50, and $>$50 keV, respectively. The solid black curve gives the sum. (c) The evolution of the average photon energy (black solid curve) and its standard deviation (black dashed curve) for the case of the central seed photons. The evolution for the uniform seed photons is similar (except that it starts at $t=0$). Time is measured in units of the light travel time across the source radius, $R/c$. The optical depth covered by a photon during that time equals $\tau t/(R/c)$.
}
\label{up_time}
\end{figure}

\begin{figure}
\centerline{\includegraphics[height=6.7cm]{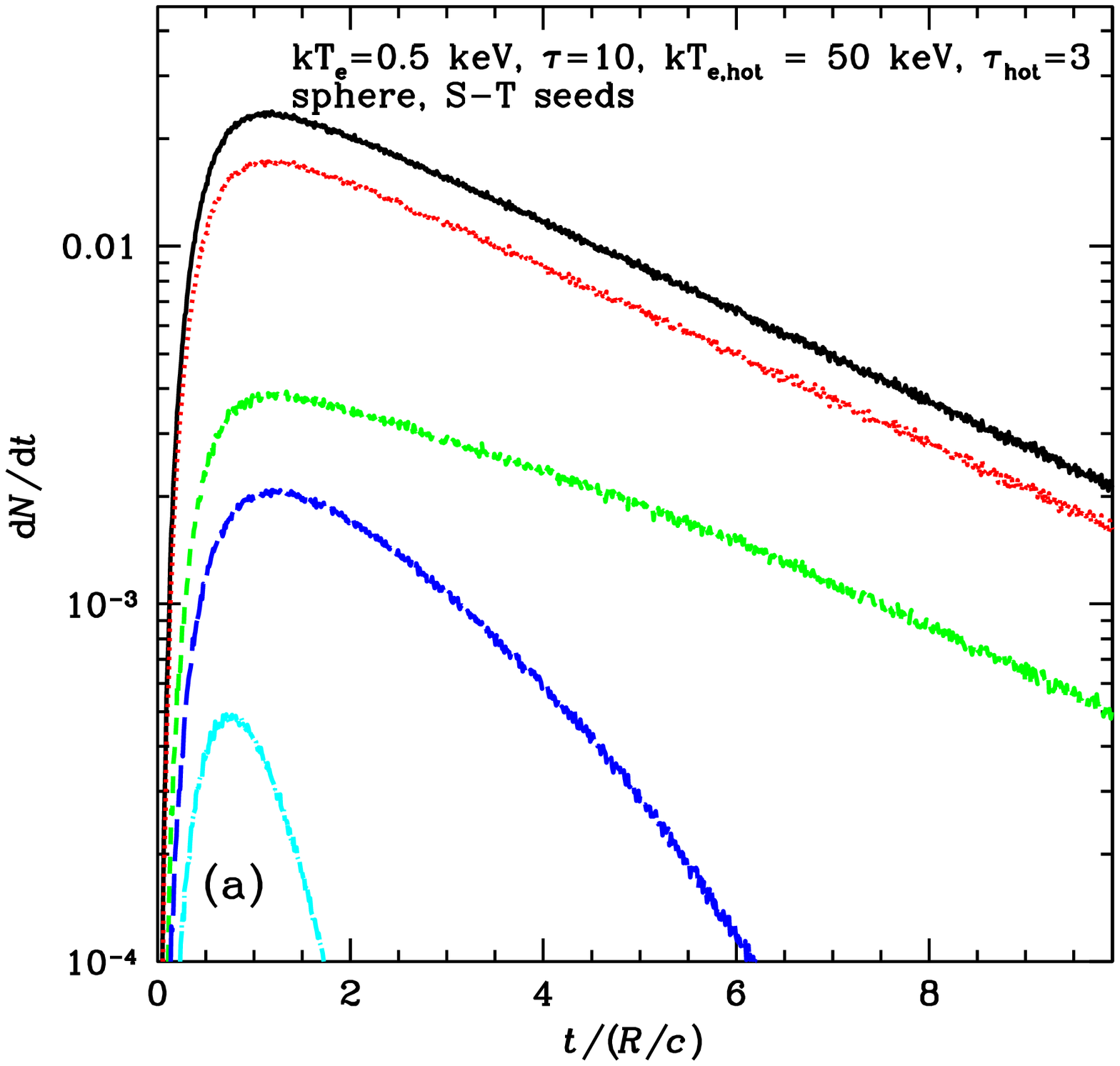}}
\centerline{\includegraphics[height=6.7cm]{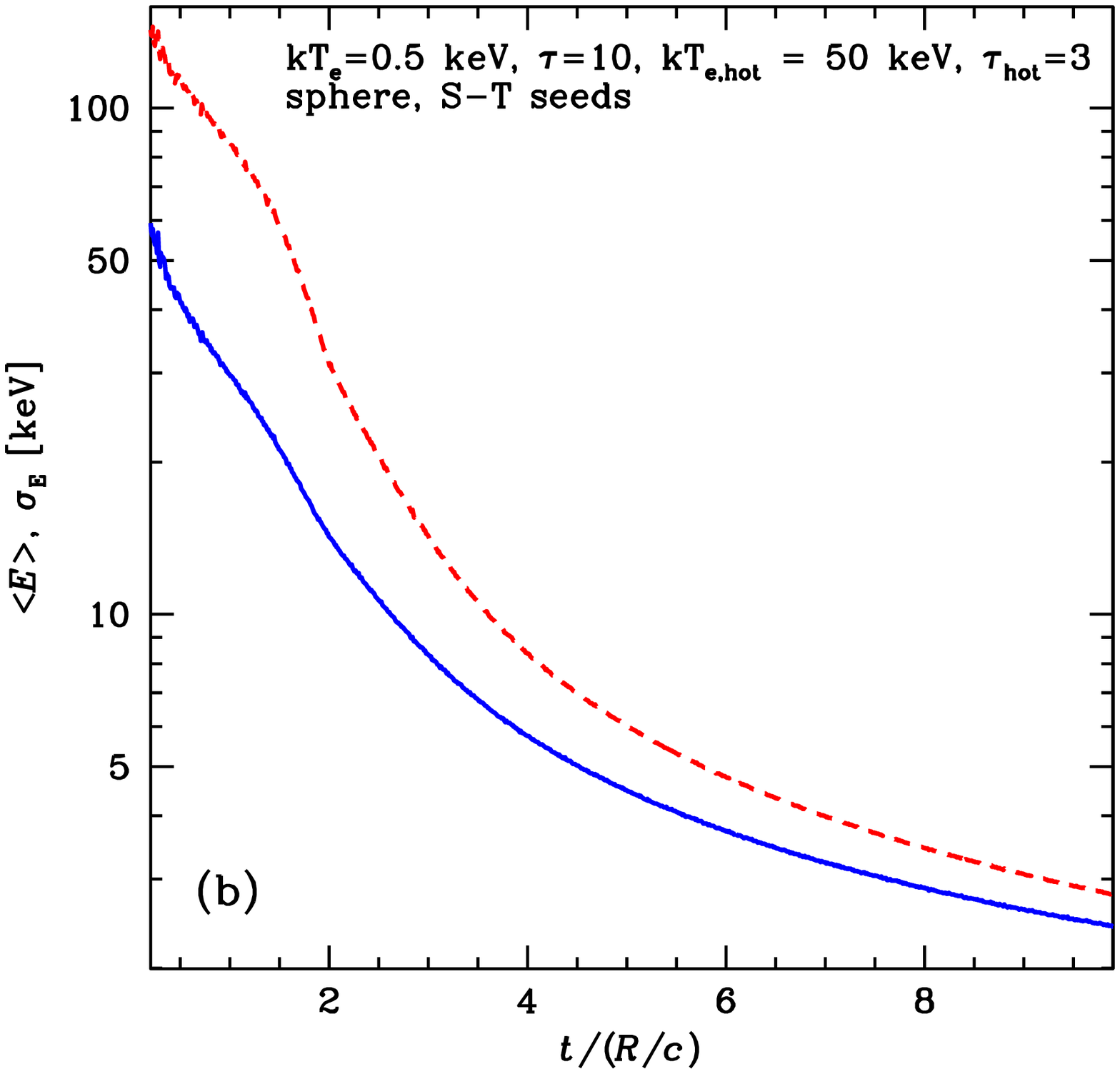}}
\caption{Timing properties of down-scattering (modelled by \compton) in a cold plasma with $kT_{\rm e}=0.5$\,keV, $\tau =10$, of an initial hard spectrum from thermal Comptonization with $kT_{\rm e, hot}=50$\,keV, $\tau_{\rm hot} =3$, $kT_{\rm bb}=0.1$\,keV (modelled by \compps) distributed sinusoidally in the sphere. The spectra for this case are shown in Fig.\,\ref{down_ISM}. Only scattered photons are now taken into account. (a) The distributions of the arrival times for central seed photons divided by the energy range, with the red dots, green short dashes, blue long dashes and cyan dot-dashed curve corresponding to the ranges $<$2, 2--10, 10--50, and $>$50 keV, respectively. The solid black curve gives the sum. (b) The evolution of the average photon energy (blue solid curve) and its standard deviation (red dashed curve). Time is measured in units of the light travel time across the source radius, $R/c$. The optical depth covered by a photon during that time equals $\tau t/(R/c)$.
}
\label{down_ISM_time}
\end{figure}

Timing properties of thermal Comptonization have been considered, e.g., by \citet*{lightman81}, \citet{kylafis87}, \citet*{kazanas97}, \citet{nowak99}, \citet*{maccarone00}, \citet{poutanen01}, \citet{zmg10}, \citet{papadakis16}, \citet{chainakun19a}, \citet{chainakun19}, \citet*{zhang19}. Compton scattering can affect timing properties of accreting sources for either up or down-scattering. 

Compton scattering in an optically thick and relatively cold plasma cloud surrounding the central X-ray source appears to be the only viable explanation of the lack of X-ray variability at frequencies $\ga 1$\,Hz in the X-ray binary Cyg X-3 (e.g., \citealt*{axelsson09}), as proposed by \citet{zmg10}. In that case, the required size of the scattering cold plasma is rather large, with a radius of $\sim\! 10^9\,{\rm cm}\approx 10^4 R_{\rm g}$, where $R_{\rm g}\equiv GM/c^2$, and $M$ is the compact object mass. 

On the other hand, X-ray sources in most of accreting black-hole binaries and AGNs appear to have sizes of $\sim\! 10^1$--$10^3 R_{\rm g}$ (e.g., \citealt{dgk07,demarco13,demarco15,chartas16, bernardini16,dzielak19}; \citealt*{mahmoud19}). This rules out the observed long hard X-ray lags (e.g., in Cyg X-1, \citealt*{miyamoto89,kotov01}) to be due to time lag between subsequent scatterings in a hot plasma. Furthermore, the lags decrease with the increasing Fourier frequency and the autocorrelation function becomes narrower with the increasing photon energy, which both rule out Compton scattering in a large corona \citep{nowak99,maccarone00}. Still, Compton scattering can be detectable in timing properties on short (in units of $R_{\rm g}/c$) time scales, in particular in AGNs (e.g., \citealt{chainakun19}). Expected characteristic time scales are, for example, $20 R_{\rm g}/c\approx 1$\,ms at $10\msun$ and $2 R_{\rm g}/c\approx 10^3$\,s at $10^8\msun$. 

We point out that a variable flux of seed photons irradiating a hot plasma cloud will affect its cooling rate. This will change the plasma temperature, the slope of the X-ray spectrum, and consequently, the time lags and correlations between light curves at different energies, see, e.g., a discussion in \citet{poutanen01}. In addition, the heating rate can vary as well. Here, we neglect these effects, and study only static Compton clouds. This requires that the variability amplitude of the input seed photons is small enough not to significantly affect the slope of the X-ray spectrum. On the other hand, in the case of a cold plasma down-scattering X-rays the plasma temperature has only a slight effect on this process, and the effect of the response of the plasma to irradiation is usually not important.

If the intrinsic variability of the seed photon flux is $S(t)$, the light curve observed after Compton scattering, $F(t)$, is a convolution,
\begin{equation}
F(t)= \int_{t_0}^\infty S(t-t') G(t') {\rm d}t',
\label{convolution}
\end{equation}
where $G(t)$ is the Green's (response) function for the problem, i.e., the distribution of the time delays of the emission due to Compton scattering. The Fourier transform of $F(t)$, ${\cal F}$, is then the product of the individual transforms, i.e., 
\begin{align}
&{\cal F}(f)={\cal S}(f) {\cal G}(f), \label{product}\\
&{\cal G}(f)=\int_{-\infty}^\infty G(t) {\rm e}^{-2\upi{\rm i}f t} {\rm d}t, \quad {\cal S}(f)=\int_{-\infty}^\infty S(t) {\rm e}^{-2\upi{\rm i}f t} {\rm d}t, \label{fourier}
\end{align}
where $f$ is the frequency. The power spectrum at $f$ is given by $\left|{\cal F}(f)\right|^2$. If the original signal is sinusoidal, its amplitude is reduced by $\left|{\cal G}(f)\right|$. In general, damping due to scattering of an intrinsic variability can be calculated either using the convolution of the light curves, equation (\ref{convolution}), or by integrating the power spectrum. The intrinsic power spectrum at $f$ is damped by $\left|{\cal G}(f)\right|^2$.

We then show an example of the timing properties for Compton upscattering of soft photons by hot plasma, with the parameters for which the spectra are shown in Fig.\,\ref{seed_geo}. Fig.\,\ref{up_time}(a) shows the distributions of the photon arrival times, ${\rm d}N/{\rm d}t \equiv G(t)$, for the cases of central and uniform distributions of the seed photons, measured from the centre of the sphere at a plane tangent to the sphere (which is equivalent to that measured by a remote observer, apart from a constant time shift). Time is measured in units of the light travel time across the source radius, $R/c$. For a known size of the source, $R$, the plotted time can be easily converted into the physical time. We see that the shapes of the two distributions are very similar apart from a difference at times comparable to the light travel time across the source. As noted by \citet{zmg10}, there is a difference between the distributions of the photon arrival times at the sphere boundary and at a remote observer, with the arrival times measured by the latter shifted up by $\Delta t=(R/c)(1-\cos\alpha)$, where $\alpha$ is the angle at which a photon leaves the sphere with respect to the radial direction. This effect is taken into account in the code, and it is responsible for the wiggle seen close to $R/c$ in Fig.\,\ref{up_time}(a) for the case of the central seed photons. We see that the response function is very asymmetric, and it has an extended exponential tail. Thus, its shape is far from the Gaussian, which form has sometimes been assumed.

Fig.\,\ref{up_time}(b) shows ${\rm d}N/{\rm d}t$ for selected photon energy ranges. We see shifts to longer delays with the increasing energy range, expected for upscattering. This increase then results in time lags between the energy bands. Thus, Compton upscattering can naturally produce the asymmetric shape of the responses, with responses in higher-energy bands dominating at later times, cf.\ \citet{chainakun19}. Fig.\,\ref{up_time}(c) shows the corresponding evolution of the average photon energy, $\langle E\rangle$, and its standard deviation, $\sigma_{\rm E}$, for the case of central seed photons. The average energy increases after each scattering until it saturates close to the average energy of the Wien peak at several tens of $R/c$. The standard deviation of the photon energy is initially greater than the average, and it becomes relatively low in the Wien peak. The analogous curves for the uniform seed photon case are similar.

Then, we show an example of the timing properties for Compton down-scattering of an initial hard spectrum by a cold plasma. Fig.\,\ref{down_ISM_time}(a) shows the distributions of the photon arrival times. It also shows ${\rm d}N/{\rm d}t$ for selected photon energy ranges. We see shifts to longer average delays with the decreasing energy range, as expected for down-scattering. Fig.\,\ref{down_ISM_time}(b) shows the corresponding evolution of the average photon energy and its standard deviation, with both $\langle E\rangle$ and $\sigma_{\rm E}$ decreasing now. The seed photon distribution is from thermal Comptonization by a hot plasma, as shown in Fig.\,\ref{down_ISM}. 

\citet{zmg10} calculated the damping effect of down-scattering in the case of Cyg X-3 using the entire range of the photon energies, while the Green's functions depend on that range. However, as shown in Fig.\,\ref{down_ISM_time}(a), this dependence is mild for photon energies $\la$10 keV, which range dominates power spectra currently measured in cosmic sources, and in Cyg X-3 in particular. Thus, their results remain valid.

\section{Conclusions}
\label{conclusions}

We have studied modifications to the kinetic-equation solution of \citetalias{st80}, which was given for nonrelativistic thermal Comptonization in a spherical source with a large Thomson optical depth. Our goal was to achieve a high accuracy of the solution in the mildly relativistic regime and $\tau\ga 1$. Our modifications to the kinetic equation are phenomenological, and based on comparison with Monte Carlo results. Our modified kinetic equation, and the corresponding code, \thcomp, are described in Section \ref{thcomp}. We also have developed a Monte Carlo code for thermal Comptonization, \compton, capable of treating various geometries and electron distributions and yielding both spectral and timing properties of the scattering cloud. For the sake of simplicity, we have considered only the spherical source geometry.

We have then compared the results of our kinetic equation with the Monte Carlo results as well as with the results of the iterative scattering method of \citet{ps96} as implemented in the code \compps. We have found (Section \ref{up}) almost perfect agreement between the Monte Carlo results and those of \compps for Compton upscattering of soft blackbody photons, with some small differences due to the effect of the finite number of iteration in the latter. Then, our kinetic-equation solution, \thcomp, agrees very well with both \compton and \compps for $kT_{\rm e}\la 300$\,keV and $\tau\ga 1.6$. The detailed parameter range in which \thcomp gives accurate results is shown in Fig.\,\ref{thcomp_limits}.

We have also compared \compton to the solution of \citet{titarchuk94}, \citet{titarchuk95}, \citet{hua95}, as given in the \xspec code \comptt. In all of the tested cases, \comptt yields substantially softer spectra than those from either \compton or \compps in the case of uniform distribution of seed photons in a sphere, see Fig.\,\ref{comptt}. Using either central or sinusoidal seed-photon distribution worsens the agreement.

The code \thcomp is provided as a convolution function, and can be applied to any input seed-photon distribution (which is also the case for \compton). We thus have tested Compton down-scattering by cold plasma of hard photon distributions extending up to $E\sim m_{\rm e}c^2$ (Section \ref{down}). We have found that \thcomp works well in this case as well. 

Finally, we have studied timing properties of Comptonization using \compton (Section \ref{timing}). We present some example distributions of the photon arrival times (representing the Green's function for the problem) and the evolution of the average photon energy for both up and down-scattering. While the Green's function at a given time lag is averaged over the photon energies, the average photon energy strongly changes with the lag. This effect should be taken into account in comparing with data. 

\section*{Acknowledgments}

We thank Barbara De Marco for valuable discussions, and the referee for valuable comments. This research has been supported in part by the Polish National Science Centre grant 2015/18/A/ST9/00746 and by the grant 14.W03.31.0021 of the Ministry of Science and Higher Education of the Russian Federation. 

\bibliographystyle{mn2e}
\bibliography{references}

\label{lastpage}

\end{document}